\documentclass[12pt]{article}
\usepackage{a4wide,amssymb,cite}
\pdfoutput=1

\usepackage{a4wide,amssymb,cite,graphicx}
\usepackage{epsfig}
\usepackage[usenames,dvipsnames]{color}
\usepackage{slashed}
\newcommand{\be}{\begin{equation}}
\newcommand{\ee}{\end{equation}}
\newcommand{\bea}{\begin{eqnarray}}
\newcommand{\eea}{\end{eqnarray}}

\def\circa#1{\,\raise.3ex\hbox{$#1$\kern-.75em\lower1ex\hbox{$\sim$}}\,}

\begin{document}

\begin{titlepage}
%
%


%

\begin{centering}
\vspace{1cm}
{\Large {\bf On thermal production of self-interacting dark matter}} \\

\vspace{1.5cm}

{\bf Soo-Min Choi, Yoo-Jin Kang and  Hyun Min Lee}
\vspace{.5cm}

{\it Department of Physics, Chung-Ang University, Seoul 06974, Korea.} 
\\

\end{centering}
\vspace{2cm}

\begin{abstract}
\noindent
We consider thermal production mechanisms of self-interacting dark matter in models with  gauged $Z_3$ symmetry.  A complex scalar dark matter is stabilized by the $Z_3$, that is the remnant of a local dark $U(1)_d$.  Light dark matter with large self-interaction can be produced from thermal freeze-out in the presence of SM-annihilation, SIMP and/or forbidden channels. We show that dark photon and/or dark Higgs should be relatively light for unitarity and then assist the thermal freeze-out. We identify the constraints on the parameter space of dark matter self-interaction and mass in cases that one or some of the channels are important in determining the relic density.

\end{abstract}

\vspace{3cm}

\end{titlepage}

\section{Introduction}

Dark matter(DM) is the dominant component of matter in the Universe, and evidences for dark matter from galaxy rotation curves, gravitational lensing, and Cosmic Microwave Background have been getting more diverse and precise. For instance, the averaged relic density of dark matter is inferred by Planck data to be $\Omega_{\rm DM} h^2=0.1198\pm 0.0015$ \cite{planck}. Weakly Interacting Massive Particles (WIMP) have been a well-motivated candidate for dark matter carrying weak interaction and weak-scale mass. The freeze-out mechanism for producing dark matter in the early Universe sets the annihilation cross section of dark matter to $\langle\sigma v\rangle_{\rm ann}\sim{\rm pb}\cdot c$, which enables us to take different approaches to test the WIMP scenario. 
Therefore, there have been a lot of complementary efforts \footnote{See, for instance, Ref.~\cite{kang} for the interplay between collider excesses and dark matter detection in mediator models. } for discovering WIMP dark matter, from direct detection, indirect detection and collider searches, but there have been null results until now. In particular, various direct detection experiments such as XENON100 \cite{xenon100} as well as LUX \cite{lux2013,lux2015} and PandaX-II \cite{pandax} have quite much constrained the WIMP-nucleon scattering cross section at the order of sub-zepto barn ($10^{-46}\,{\rm cm}^2$).

On the other hand, dark matter is assumed to be collisionless, namely, carry no self-interactions, in Standard Cosmology, the so called $\Lambda$CDM. But, the numerical simulation with collisionless dark matter would lead to cuspy DM profiles that are not consistent with observed galaxies (core-cusp problem), as well as too many sub-halos (missing satellite problem) and too large masses for sub-halos (too-big-to-fail problem). These are the so called small-scale problems at galaxy scales \cite{smallscale1,smallscale2}. Although the inclusion of baryons and supernova feedback in simulations might resolve such tensions in massive galaxies \cite{baryon}, the small-scale problems persist in the lowest mass galaxies where the discrepancy exists \cite{review}. Therefore, small-scale problems may call for strong dark matter self-interactions, leading to $\sigma_{\rm self}/m_{\rm DM}=0.1-10\,{\rm cm^2/g}$ or $\sigma_{\rm self}/m_{\rm DM}\sim {\rm barn}$ for $m_{\rm DM}\sim 1\,{\rm GeV}$.

Strongly Interacting Massive Particles(SIMP) \cite{simp1a} have recently drawn attention, due to the fact that the thermal freeze-out with $3\rightarrow 2$ annihilation \cite{carlson} allows for a large self-scattering of light dark matter. It is the Boltzmann suppression factor associated with an extra dark matter particle in the $3\rightarrow 2$ process that naturally generates a hierarchy between the thermal annihilation cross section of about pico-barn and the self-scattering cross section of about barn \cite{simp-review}. However, the Boltzmann suppression factor is more or less fixed to $e^{-x_f}$ with $x_f=m_{\rm DM}/T_f$ at freeze-out temperature $T_f$. Therefore, the relic density condition needs a relatively large self-interaction of SIMP dark matter, that is on the verge of violating unitarity or perturbativity and is in a tension with the bounds from Bullet cluster and halo shapes in most of the parameter space. There have been quite a few works in the literature for proposing concrete models to realize the SIMP \cite{simp1b,z3global,simp2,z3dm,simp-review,z5simp,simpmore} and its variations \cite{elder,cannibal,glueball}. There have been growing interests in detecting the light dark matter of sub-GeV scale from direct detection \cite{xenon10,essig,superconductor,helium,graphene} and cosmic rays \cite{cosmicrays}.

In this article, we consider a complex scalar dark matter with gauged $Z_3$ symmetry in light of self-interacting dark matter \cite{z3dm}. The $Z_3$ coming from the spontaneous breaking of a local $U(1)_d$ stabilizes dark matter while the resultant dark photon and dark Higgs can contribute to the determination of the relic density. In particular, the semi-annihilation of dark matter \cite{semiann} into heavier dark photon or dark Higgs, the so called forbidden channels \cite{forbidden1,forbidden2}, can be suppressed by a Boltzmann factor $e^{-\Delta_i x_f}$ , with $\Delta_i=(m_i-m_{\rm DM})/m_{\rm DM}$, where $m_{\rm DM}$ is the DM mass and $m_i$ is the mass of dark photon or dark Higgs. Then, taking $\Delta_i\lesssim 1$, it is possible to accommodate a smaller self-interaction of dark matter being compatible with the relic density, as compared to the SIMP case.  Interestingly, the cubic self-coupling of dark matter in our model contributes to both SIMP and forbidden channels.  

Furthermore, there exists a standard $2\rightarrow 2$ annihilation of light dark matter into a pair of SM particles, in the presence of a $Z'$ portal coupling \cite{hambye}. In this case, the smallness of the standard $2\rightarrow 2$ annihilation (SM-annihilation) could be attributed to the smallness of the gauge kinetic mixing between $U(1)_d$ and hypercharge gauge group. 
We provide the general discussion on the thermal production of self-interacting dark matter in our model, in cases that one or some of SM-annihilation, SIMP and forbidden channels are relevant.

The paper is organized as follows. 
We begin with a review on the model with gauged $Z_3$ symmetry for dark matter and discuss  the resultant mass spectrum and vacuum stability of the model. 
Then, we make general discussion on self-scattering and Boltzmann equation and kinetic equilibrium condition in the model.  Next we study three mechanisms for thermal production, namely, SM-annihilation, SIMP and forbidden channels, in the presence of non-decoupled dark photon and dark Higgs, and discuss the constraints on the model from the relic density, self-scattering and various collider searches for light dark matter.   Then, conclusions are drawn. 
There is one appendix dealing with the $2\rightarrow 2$ forbidden channels involving dark photon and dark Higgs in our model.

\section{Model for self-interacting dark matter}

We consider dark matter as a complex scalar $\chi$ having a charge $q_\chi=+1$ under the dark local $U(1)_d$ symmetry, which is spontaneously broken to $Z_3$ by the VEV of another complex scalar $\phi$ with charge $q_\phi=+3$.  Thus, the remaining discrete $Z_3$ symmetry \footnote{See Ref.~\cite{z3global,z3global2} for the discussion on global $Z_3$ symmetry for dark matter.} ensures the stability of scalar dark matter $\chi$ \cite{ko,z3dm}. 

The Lagrangian for SM singlet scalars, $\chi, \phi$, and the SM Higgs doublet $H$, is given \cite{ko,z3dm} by
\bea
{\cal L}=-\frac{1}{4}V_{\mu\nu} V^{\mu\nu}-\frac{1}{2}\sin\xi \,V_{\mu\nu} B^{\mu\nu}+ |D_\mu\phi|^2 + |D_\mu \chi|^2 + |D_\mu H|^2 - V(\phi,\chi,H)
\eea
where the field strength tensor for dark photon is $V_{\mu\nu}=\partial_\mu V_\nu-\partial_\nu V_\mu$, and covariant derivatives are $D_\mu \phi=(\partial_\mu - iq_\phi g_d V_\mu)\phi$,  
$D_\mu\chi = (\partial_\mu - i q_\chi g_d V_\mu)\chi$, with $g_d$ being dark gauge coupling, and $D_\mu H= (\partial_\mu - ig' Y_H B_\mu -\frac{1}{2} i g T^a W^a_\mu) H$, and the gauge kinetic mixing between dark photon  $V_\mu$ and hypercharge gauge boson $B_\mu$ is introduced by $\sin\xi$.  Then, the dark photon communicates between dark matter and the SM particles through the gauge kinetic mixing. 
Here, the scalar potential is $V(\phi,\chi,H)= V_{\rm DM}+V_{\rm SM}$
with
\bea
V_{\rm DM}&=& -m^2_\phi |\phi|^2 + m^2_{\chi} |\chi|^2 +\lambda_\phi |\phi|^4 +\lambda_\chi |\chi|^4 + \lambda_{\phi \chi} |\phi|^2 |\chi|^2\nonumber \\
&&+\bigg(\frac{\sqrt{2}}{3!}\,\kappa \phi^\dagger \chi^3+{\rm h.c.} \bigg)+ \lambda_{\phi H}|\phi|^2 |H|^2 + \lambda_{\chi H} |\chi|^2 |H|^2, \\
V_{\rm SM}&=&  -m^2_H |H|^2 +\lambda_H |H|^4. 
\eea
We note that the presence of a dark Higgs $\phi$ allows a triple coupling for $\chi$ after the $U(1)_d$ is spontaneously broken. Therefore, the corresponding $\kappa$ coupling leads to SIMP processes as well as (forbidden) semi-annihilation processes of dark matter, which will be relevant for the later discussion.

\subsection{Mass spectrum}

For a nonzero VEV of dark Higgs field with $\langle \phi\rangle=\frac{1}{\sqrt{2}}v_d$, the $U(1)_d$ symmetry is broken to a discrete subgroup $Z_3$ and dark photon gets massive and can mix with photon and $Z$-boson. 
After expanding the dark Higgs as $\phi=\frac{1}{\sqrt{2}}(v_d+h_d)$ and taking the SM Higgs doublet to be $H^T=\frac{1}{\sqrt{2}}(0,v_{\rm ew}+h)$, the dark Higgs can mix with the SM Higgs by Higgs-portal interaction, $\lambda_{\phi H}$. 
Then, the SM and dark Higgs bosons are mixed  \cite{z3dm} by
\be
\left( \begin{array}{c} h_1 \\ h_2 \end{array} \right)=  \left( \begin{array}{cc} \cos\theta & -\sin\theta \\  \sin\theta & \cos\theta \end{array} \right) \left( \begin{array}{c} h_d \\ h \end{array} \right)
\ee
where $h_1,h_2$ are mass eigenstates. 
The mass eigenvalues of Higgs-like states are
\be
m^2_{h_1,h_2}=\lambda_\phi v^{ 2}_d+\lambda_H v^2_{\rm ew} \mp \sqrt{(\lambda_\phi v^{ 2}_d-\lambda_H v^2_{\rm ew})^2+ \lambda^2_{\phi H}v^{2}_d v^2_{\rm ew} }
\ee
and the mixing angle is 
\be
\tan 2\theta = \frac{\lambda_{\phi H} v_d v_{\rm ew}}{\lambda_H v^2_{\rm ew}-\lambda_\phi v^{2}_d}. 
\ee
On the other hand, the effective mass of dark matter is given by $m^2_{\chi,{\rm eff}}=m^2_{\chi}+\frac{1}{2}\lambda_{\lambda\chi} v^2_d+\frac{1}{2}\lambda_{\chi H} v^2_{\rm ew}$, but we can absorb the contributions from symmetry breaking into the bare mass of dark matter.
The details of interaction terms for dark/SM Higgses and dark photon can be found in Ref.~\cite{z3dm}.

Moreover, the mass eigenvalues of $Z$-boson and dark photon are
\bea
m^2_{1,2}= \frac{1}{2}\left[m^2_Z (1+s^2_W t^2_\xi)+m^2_V/c^2_\xi\pm \sqrt{(m^2_Z(1+s^2_W t^2_\xi)+m^2_V/c^2_\xi)^2- 4m^2_Z m^2_V /c^2_\xi} \,\right]
\eea
where $m^2_Z=\frac{1}{4}(g^2+g^{\prime 2}) v^2$ and $m^2_V=9g^2_d v^2_d$, and the mixing angle between $Z$-boson and dark photon is given by
\be
\tan 2\zeta =\frac{m^2_Z s_W \sin 2\xi}{m^2_V-m^2_Z(c^2_\xi-s^2_W s^2_\xi )}.
\ee
Then, taking $\zeta\simeq -s_W  \xi$ for $m_V\ll m_Z$, we obtain the current interactions for dark photon as
\bea
{\cal L}_{Z_2,{\rm int}}\approx Z_{2\mu}  \Big(- e \varepsilon J^\mu_{\rm EM}+g_d J^\mu_d \Big)
\eea
 where $\varepsilon\equiv c_W \xi$, and $J^\mu_{\rm EM}$ and $J^\mu_d$ are electromagnetic, neutral and dark currents, respectively.    In this case, we get $m_2\approx 3 g_d v_d\equiv m_{Z'}$.
 See the appendix A of Ref.~\cite{z3dm} for the details.

\subsection{Vacuum stability}

The absolute vacuum stability requires the potential to be bounded from below, meaning that $V>0$ for large field values away from the local minimum with $V=0$. 
In this section, for simplicity, we focus on the vacuum stability in the hidden sector with dark Higgs and dark matter scalars only. Although the mixing quartic couplings with the SM Higgs can affect our discussion too, they can be safely ignored, when they take positive or small values as compared to couplings in the hidden sector. 

Taking $\phi=\frac{1}{\sqrt{2}}\,\alpha$ and $\chi=\frac{1}{\sqrt{2}}\,\beta \, e^{i\gamma}$ for large field values, the vacuum stability is determined by the quartic couplings in $V_{\rm DM}$, which becomes
\bea
V_{\rm DM}= \frac{1}{4}\lambda_\phi \alpha^4 +\frac{1}{4}\lambda_\chi \beta^4+\frac{1}{4}\lambda_{\phi\chi} \alpha^2\beta^2 +\frac{\sqrt{2}}{12}\,\kappa\alpha\beta^3\cos(3\gamma).
\eea
After minimizing the potential for $\gamma$ along any field values with $\alpha\neq 0$ and $\beta\neq 0$, the above hidden sector potential becomes
\bea
V_{\rm DM}= \frac{1}{4}\lambda_\phi \alpha^4 +\frac{1}{4}\lambda_\chi \beta^4+\frac{1}{4}\lambda_{\phi\chi} \alpha^2\beta^2 -\frac{\sqrt{2}}{12}\,|\kappa||\alpha||\beta|^3.
\eea
Therefore, the vacuum stability conditions are given by
\bea
\lambda_\phi>0,\quad \lambda_\chi>0,  \label{1vsc}
\eea
and
\bea
f(X_{\rm min})>0 \label{3vsc0}
\eea
with
\bea
f(X)=\frac{1}{4}\lambda_\phi X^4+\frac{1}{4}\lambda_{\phi\chi} X^2-\frac{\sqrt{2}}{12}\,|\kappa| X+\frac{1}{4}\lambda_\chi,
\eea
where $X_{\rm min}$ is the global minimum satisfying $f'(X_{\rm min})=0$.
Then, solving $f'(X_{\rm min})=0$,  we obtain the third condition (\ref{3vsc0}) as
\be
\lambda_\chi> -\frac{1}{2}\lambda_{\phi\chi} X^2_{\rm min}+\frac{\sqrt{2}}{4}|\kappa| X_{\rm min}  \label{3vsc0}
\ee 
with
\be
X_{\rm min}= \left\{ \begin{array}{cc} (P+\sqrt{P^2+Q^3})^{1/3}+(P-\sqrt{P^2+Q^3})^{1/3},\quad D>0, \\
2\sqrt{-Q}\cos\Big(\frac{1}{3}\arccos\Big(\frac{P}{\sqrt{-Q^3}}\Big)\Big),\quad D<0, \end{array}\right.
\ee
where $D\equiv P^2+Q^3$ with $P\equiv \frac{\sqrt{2}|\kappa|}{24\lambda_\phi}$ and $Q\equiv \frac{\lambda_{\phi\chi}}{6\lambda_\phi}$.

For instance, for $\kappa=0$, the third vacuum stability condition (\ref{3vsc0}) becomes trivial for $\lambda_{\phi\chi}>0$, while it is given 
by $4\lambda_\phi\lambda_\chi-\lambda^2_{\phi\chi}>0$ for $\lambda_{\phi\chi}<0$, which is the standard result for two scalar fields with a mixing quartic coupling. 
On the other hand, for $\lambda_{\phi\chi}=0$, the third vacuum stability condition (\ref{3vsc0}) becomes $192\lambda_\phi \lambda^3_\chi -\kappa^4>0$.

The vacuum stability condition equivalent to eq.~(\ref{3vsc0}) can be also derived by the condition that there is no real solution to the quartic polynominal $f(X)$, in a more explicit form \cite{kannike},
\bea
\lambda_{\phi\chi}+2\sqrt{\lambda_\phi\lambda_\chi}>0  \label{3vsc1}
\eea
and
\bea
36\lambda_\chi (\lambda^2_{\phi\chi}-4\lambda_\phi \lambda_\chi)^2>2\lambda_{\phi\chi}\kappa^2 (\lambda_{\phi\chi}^2-36\lambda_\phi \lambda_\chi)+3 \lambda_\phi \kappa^4.
\label{3vsc}
\eea
Then, the conditions, (\ref{3vsc0}), (\ref{3vsc1}) and (\ref{3vsc}), turn out to be equivalent.

For negative Higgs mixing quartic couplings with $\lambda_{\phi H}<0$ and $\lambda_{\chi H}<0$, there are corresponding vacuum stability conditions for them too. But, in the later analysis, we assume $\lambda_{\phi H}, \lambda_{\chi H}$ to be positive if they are nonzero, so there is no extra conditions for vacuum stability. The general discussion on the vacuum stability conditions with arbitrary $\lambda_{\phi H}$ and $\lambda_{\chi H}$ are given in Ref.~\cite{kannike}.

For the later sections, we will impose the vacuum stability conditions, (\ref{1vsc}) and (\ref{3vsc}), for the consistency of the vacuum breaking the $U(1)_d$.

\section{Dynamics of self-interacting dark matter}

We discuss the self-scattering of dark matter and resulting unitarity bounds and present the general Boltzmann equation for the early Universe in our model. Then, we comment on the kinetic equilibrium condition for dark matter and the elastic scattering between light dark matter and electron.

\subsection{Dark matter self-scattering and unitarity bounds}

The squared amplitude for  the $\chi\chi\rightarrow \chi\chi$ self-scattering is given \cite{z3dm} by 
\be
|{\cal M}_{\rm \chi\chi}|^2=2\left(2\lambda_\chi+ 3R^2+\frac{4g^2_d m^2_\chi}{m^2_{Z'}} -\frac{\lambda_{\phi\chi}^2 m_{Z'}^2}{9g^2_d m^2_{h_1} }  \right)^2  \label{chichi}
\ee
with $R\equiv \sqrt{2}\kappa v_d/(6m_\chi)$.
On the other hand, the squared amplitude for the $\chi\chi^*\rightarrow \chi\chi^*$ self-scattering is given \cite{z3dm} by
\bea
|{\cal M}_{\rm \chi\chi^*}|^2=4\left(2\lambda_\chi- 9 R^2-\frac{2g^2_dm^2_\chi}{m^2_{Z'}}+ \frac{\lambda_{\phi\chi}^2 (-2 m_\chi^2 + m_{h_1}^2) m_{Z'}^2}{9g_d^2 (4m_\chi^2 - m_{h_1}^2)m_{h_1}^2} \right)^2.  \label{chichistar}
\eea
Therefore, in the non-relativistic limit for dark matter,  the effective scattering cross section, $\sigma_{\rm self}\equiv \frac{1}{4}(\sigma^{\chi\chi}_{\rm self}+\sigma^{\chi^*\chi^*}_{\rm self}+ \sigma^{\chi\chi^*}_{\rm self} )$ with $\sigma^{\chi^*\chi^*}_{\rm self}=\sigma^{\chi\chi}_{\rm self}$, is
\bea
\sigma_{\rm self} = \frac{1}{64\pi m^2_\chi}\left( |{\cal M}_{\rm \chi\chi}|^2+|{\cal M}_{\rm \chi\chi^*}|^2 \right).
\eea

The perturbativity and unitarity bounds on the DM couplings are given as follows,
\bea
\lambda_\phi, \lambda_\chi< 4\pi,  \quad \quad  |{\cal M}_{\chi\chi}|, |{\cal M}_{\chi\chi^*}| < 8\pi. 
\eea
In the later sections, we will impose the above unitarity and perturbativity conditions for the consistency of the model.

\subsection{General Boltzmann equation}

Assuming CP conservation in the dark sector, we obtain the general Boltzmann equation for dark matter number density in our model, $n_{\rm DM}=n_{\chi}+n_{\chi^*}$, with $n_\chi=n_{\chi^*}$,  as
 \bea
\frac{d n_{\rm DM}}{dt} + 3H n_{\rm DM} &=&  -\langle\sigma v^2\rangle_{3\rightarrow 2}(n^3_{\rm DM} -n^2_{\rm DM} n^{\rm eq}_{\rm DM} )  \nonumber \\
&&-\frac{1}{2} \langle\sigma v\rangle_{\chi\chi^*\rightarrow {\bar f} f} (n^2_{\rm DM}- (n^{\rm eq}_{\rm DM})^2) \nonumber \\
&&-\frac{1}{2}\langle \sigma v\rangle_{\chi\chi^*\rightarrow Z'Z'} n^2_{\rm DM}+2\langle\sigma v\rangle_{Z'Z'\rightarrow \chi\chi^*} (n^{\rm eq}_{Z'})^2 \nonumber \\
&& -\frac{1}{2}\langle\sigma v\rangle_{\chi\chi\rightarrow Z'\chi^*} n^2_{\rm DM}+\langle \sigma v\rangle_{Z'\chi^*\rightarrow \chi\chi} n^{\rm eq}_{Z'} n_{\rm DM} \nonumber \\
&&-\frac{1}{2}\langle \sigma v\rangle_{\chi\chi^*\rightarrow h_1 h_1} n^2_{\rm DM}+2\langle\sigma v\rangle_{h_1h_1\rightarrow \chi\chi^*} (n^{\rm eq}_{h_1})^2 \nonumber \\
&& -\frac{1}{2}\langle\sigma v\rangle_{\chi\chi\rightarrow h_1\chi^*} n^2_{\rm DM}+\langle \sigma v\rangle_{h_1\chi^*\rightarrow \chi\chi} n^{\rm eq}_{h_1} n_{\rm DM}.  \label{genBoltz}
 \eea
In principle, three annihilation processes, SM-annihilation, SIMP and forbidden channels can contribute equally in determining the number density of dark matter. In the next sections, we discuss the cases where one or some of annihilation processes become dominant. In particular, in order to make the model unitarity up to relatively large masses for self-interacting dark matter, it is necessary to introduce relatively light dark photon and/or dark Higgs so forbidden channels can be important too.

\subsection{Kinetic equilibrium and DM detection}

We assume that dark matter keeps in kinetic equilibrium during the freeze-out process, meaning that $n_{\rm SM}\langle\sigma v\rangle_{\chi, {\rm SM}}>H$, where $n_{\rm SM}$ is the equilibrium number density of the SM particles and $\langle\sigma v\rangle_{\chi, {\rm SM}}$ is the scattering cross section between dark matter and the SM particles in thermal bath. 
Then, we require a nonzero coupling between dark matter and the SM particles. 
To that purpose, Higgs portal or $Z'$ portal interactions in our model would be appropriate.
It turns out that Higgs portal could not be used for kinetic equilibrium of sub-GeV light dark matter, because of small Yukawa couplings. We note that there are other possibilities that can be also consistent with observations, if dark matter is in kinetic equilibrium with dark radiation, namely, dark photon in our case. If dark matter were decoupled from both the SM and dark radiation, the dark sector could undergo an epoch of heating \cite{carlson} so it would be unacceptable for structure formation.

In the later discussion on the SM-annihilating dark matter, a minimum value of the gauge kinetic mixing is needed for the correct relic density. 
Then, one has to take into account the bounds from direct detection as well as $Z'$ searches at colliders.

For $Z'$ portal interaction,  the kinetic scattering cross section for $\chi f \rightarrow \chi f $ with $f$ being the SM leptons is, in the early Universe, when leptons carries about the DM momentum, given \cite{z3dm} by
\bea
(\sigma v)_{\chi f\rightarrow \chi f}
 &=&\frac{\varepsilon^2 e^2 g^2_d m^2_\chi}{8\pi m^4_{Z'}}\,v^2 
 \label{kinsection}.
\eea
Due to cross symmetry, a nonzero kinetic scattering cross section leads to the annihilation of dark matter into $f{\bar f}$.
For sub-GeV dark matter annihilating into leptons, the X-ray and gamma-ray searches can impose strong bounds on the corresponding annihilation cross section \cite{cosmicrays}. But, in our case, as will be shown in the next section, the annihilation cross section is velocity-suppressed, so there is no limit from indirect detection \cite{z3dm}. 

Similarly, for $m_e, m_\chi, m_{Z'}\gg p\simeq m_\chi v_{\rm DM}$ at present, the DM-electron elastic scattering cross section with $Z'$-portal interaction, that is relevant for direct detection, is given \cite{z3dm} by 
\bea
\sigma_{\rm \chi e}=\frac{\varepsilon^2 e^2 g^2_d \mu^2}{\pi m^4_{Z'}}
\eea
where $\mu\equiv m_e m_\chi/(m_e+m_\chi)$ is the reduced mass of the DM-electron system. 
In the later section, we will show the parameter space that could be accessible by direct detection with semi-conductor or superconductor detectors \cite{essig,superconductor}.  The region for $m_\chi$ vs $\varepsilon$, that is consistent with the SIMP dark matter, has been also shown to be constrained by direct detection and $Z'$ searches \cite{z3dm}.

\section{Thermal freeze-out from allowed channels}

We discuss the thermal production of light dark matter from the $2\rightarrow 2$ annihilations into a pair of SM particles and the $3\rightarrow 2$ annihilations due to DM self-interactions.

\subsection{SM-annihilating dark matter}

For $m_{h_1}, m_{Z'}\gg m_\chi$, the $2\rightarrow 2$ annihilation channels are kinematically forbidden.  Furthermore, for small self-couplings of dark matter, the $3\rightarrow 2$ annihilation processes are also suppressed, namely,  $n^2_{\rm DM}\langle\sigma v^2\rangle_{3\rightarrow 2}<n_{\rm DM}\langle\sigma v\rangle_{2\rightarrow 2}$ or $n^2_{\rm DM}\langle\sigma v^2\rangle_{3\rightarrow 2}<H$.
In this case, dark matter annihilates dominantly into a pair of the SM particles. 

As a result, the Boltzmann equation (\ref{genBoltz}) is approximated to
 \bea
\frac{d n_{\rm DM}}{dt} + 3H n_{\rm DM} \approx- \langle\sigma v\rangle_{2\rightarrow 2}(n^2_{\rm DM}- (n^{\rm eq}_{\rm DM})^2) 
 \eea
where $ \langle\sigma v\rangle_{ 2\rightarrow 2}\equiv \frac{1}{2} \langle\sigma v\rangle_{\chi\chi^*\rightarrow {\bar f} f} $, which is given \cite{z3dm}, before thermal average, by 
\bea
(\sigma v )_{\chi\chi^*\rightarrow f {\bar f}}&=& \frac{\varepsilon^2 e^2 g^2_d \Big(m^2_\chi+\frac{1}{2}m^2_f\Big)}{6\pi[(4 m^2_\chi-m^2_{Z'})^2+m^2_{Z'} \Gamma^2_{Z'}] }\, \sqrt{1-\frac{m^2_f}{m^2_\chi}}\,v^2
\nonumber \\
&&+\frac{1}{4\pi}\Big(\frac{m_f}{v_{\rm ew}}\Big)^2\Big(1-\frac{m^2_f}{m^2_\chi} \Big)^{3/2}
\left|\frac{y_{h_1\chi^*\chi}}{4m^2_\chi-m^2_{h_1}}+\frac{y_{h_2\chi^*\chi}}{4m^2_\chi-m^2_{h_2}} \right|^2  \label{SMann}
\eea
with $v_{\rm ew}=246\,{\rm GeV}$, and 
\bea
y_{h_1\chi^*\chi}&\equiv& \sin\theta (\lambda_{\phi\chi}v_d\cos\theta-\lambda_{\chi H}v_{\rm ew}\sin\theta ), \\
y_{h_2\chi^*\chi}&\equiv& \cos\theta  (\lambda_{\phi\chi}v_d\sin\theta+\lambda_{\chi H}v_{\rm ew}\cos\theta ).
\eea
We note that the $Z'$-portal contribution in the first line of (\ref{SMann}) is $p$-wave suppressed while the Higgs-portal contribution in the second line of (\ref{SMann}) is suppressed by lepton Yukawa couplings. 
Thus, the model is not constrained by gamma-ray searches from the galactic center\cite{cosmicrays} or CMB constraints at recombination \cite{planck}. The SM-annihilating process with $Z'$-portal interaction is still relevant for producing a right relic density from freeze-out. 

\begin{figure}
  \begin{center}
   \includegraphics[height=0.42\textwidth]{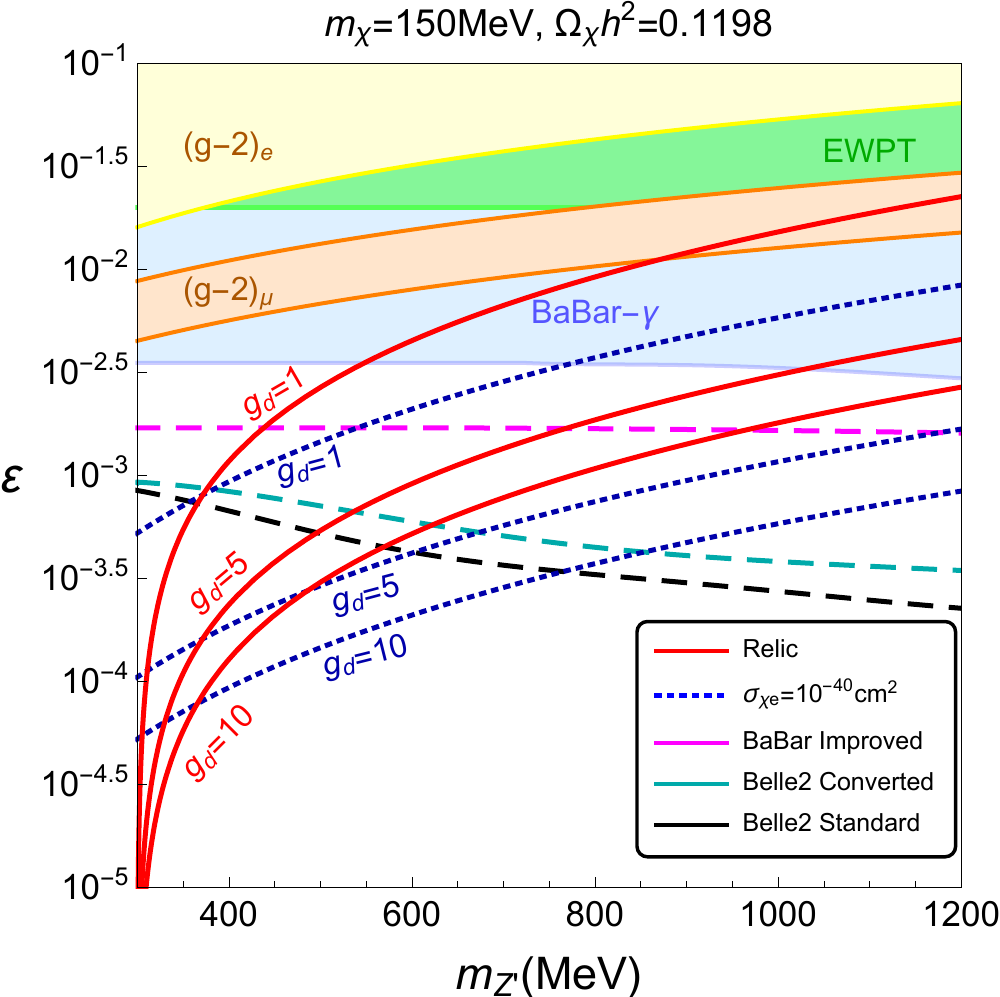} ~~~~~~~
      \includegraphics[height=0.42\textwidth]{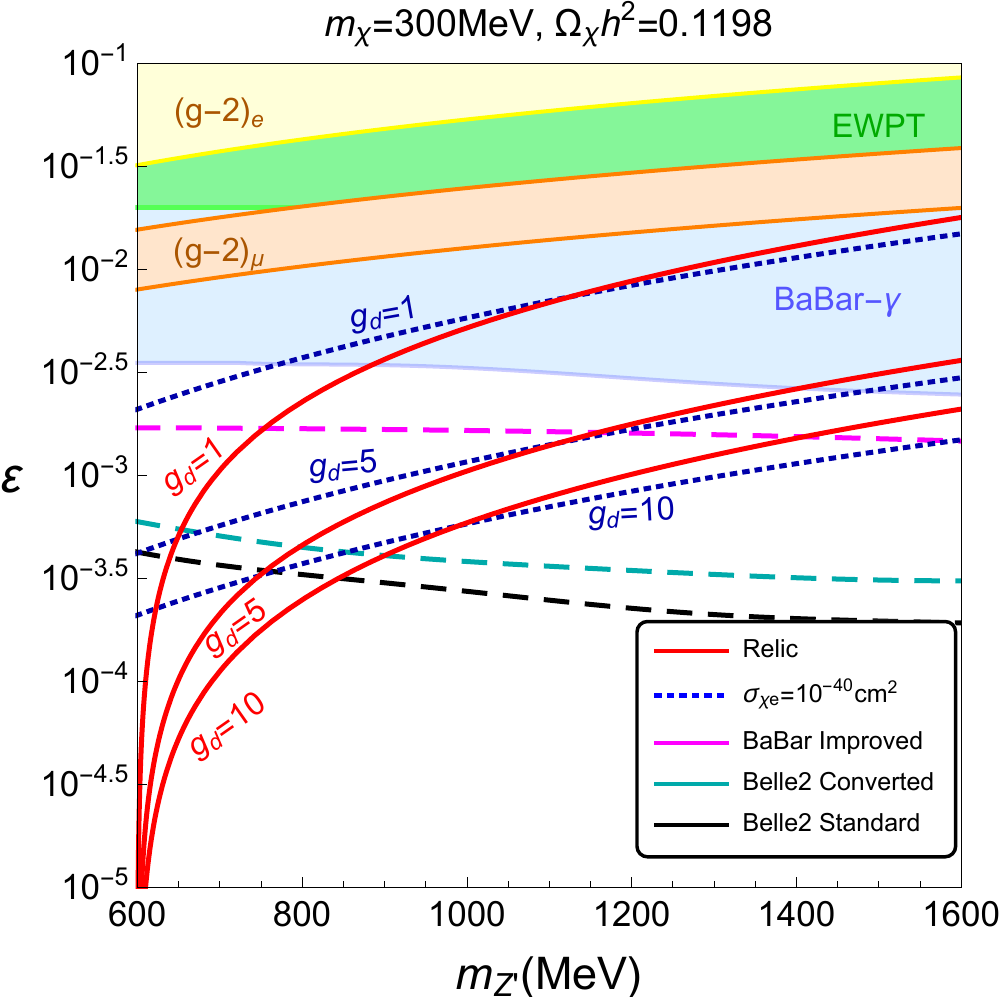}
   \end{center}
  \caption{Parameter space of $m_{Z'}$ vs $\varepsilon$, satisfying the relic density in red lines. Monophoton bounds from BaBar(improved) and Belle2(expected for converted or standard) are shown in blue region and pink, light-blue and black dashed lines. Bound from $(g-2)_e$ and favored region for $(g-2)_\mu$ are depicted in yellow and orange, respectively, while electroweak precision bound is shown in green. DM-electron scattering cross section with $\sigma_{\chi e}=10^{-40}\,{\rm cm}^2$ is shown in dotted lines too. We took DM mass to $150(300)\,{\rm MeV}$ on left (right) plots. }
  \label{sm}
\end{figure}

Consequently, for $\frac{1}{2}(\sigma v )_{\chi\chi^*\rightarrow f {\bar f}}=a+b v^2$, we get 
the relic density as
\bea
\Omega_{\rm DM} h^2 &=& \frac{8.53\times 10^{-11}{\rm GeV}^{-2}}{g^{1/2}_*  \int^\infty_{x_f} dx \, x^{-2} \langle\sigma v\rangle_{ 2\rightarrow 2} }  \nonumber  \\
&=&5.20\times 10^{-10} {\rm GeV}^{-2} \Big(\frac{g_*}{10.75} \Big)^{-1/2}\Big(\frac{x_f}{20}\Big)\left(a+\frac{3b}{x_f}\right)^{-1}.
\eea
This is the standard formula for the relic density in the case of SM annihilation, except that DM mass is taken to be sub-GeV.

In Fig.~\ref{sm}, we have shown the parameter space for $m_{Z'}$ vs $\varepsilon$ in (red) solid lines for dark matter $m_\chi=150(300)\,{\rm MeV}$ on left (right) and dark gauge coupling $g_d=1, 5, 10$, being consistent with the relic density. Electron $g-2$ limit and muon $g-2$ favored region are shown in yellow and orange colors while the bound from EWPT is given in green. Monophoton $+$ MET bounds from BaBar(improved) and Belle2(expected) \cite{monogamma} are shown in blue region and pink, light-blue and black dashed lines. The contour with elastic scattering cross section between dark matter and electron being given by $\sigma_{\chi e}=10^{-40}\,{\rm cm}^2$ are also shown in dotted lines.  We find that the region that is consistent with the relic density can be probed by semi-conductor or superconductor detectors \cite{essig,superconductor}. 

Since light dark matter annihilates into light fermions such as muons and electrons, Higgs portal interactions are Yukawa-suppressed, so they give negligible contributions to the DM annihilation. 
Nonetheless, non-negligible mixing quartic couplings $\lambda_{\chi H}$ and $\lambda_{\phi H}$, or Higgs mixing angle, would lead to additional Higgs decay modes with decay rates given by,
\bea
\Gamma(h_2\rightarrow \chi\chi^*)&=&\frac{y^2_{h_2\chi^*\chi}}{16\pi m_{h_2}}\sqrt{1-\frac{4m^2_\chi}{m^2_{h_2}}}, \\
\Gamma(h_2\rightarrow h_1 h_1)&\simeq& \frac{\lambda^2_{\phi H} v^2_{\rm ew}}{32\pi m_{h_2}}\sqrt{1-\frac{4m^2_{h_1}}{m^2_{h_2}}}.
\eea
Then, additional Higgs couplings are bounded by Higgs data of signal strengths and/or searches for Higgs invisible decays at the LHC. The combined VBF, $ZH$ and gluon fusion productions of Higgs boson at CMS lead to the bound, ${\rm BR}(h_2\rightarrow \chi\chi^*)<0.24$ at $95\%$ CL \cite{cmsHinv}, while the bounds from the VBF  \cite{atlasVBF}  and $ZH$  \cite{atlasZH}  Higgs productions at ATLAS are ${\rm BR}(h_2\rightarrow \chi\chi^*)<0.29$ and ${\rm BR}(h_2\rightarrow \chi\chi^*)<0.75$, respectively.   As a result, the bound on the Higgs invisible decay leads to $|y_{h_2 \chi^*\chi}|/v\lesssim 0.010$. 
On the other hand, the Higgs signal strength is bounded to $\mu>0.81$ at $95\%$ CL from ATLAS/CMS data combined \cite{falkowski}.   Thus, the Higgs mixing angle is bounded as $\sin\theta<0.44$, which satisfied in our case, because $\sin\theta\simeq \lambda_{\phi H} v_{\rm ew} v_d/m^2_{h_2}\lesssim 0.016$ for $m_{h_2}=125\,{\rm GeV}\gg m_{h_1}$, $v_d\sim 1\,{\rm GeV}$ and $\lambda_{\phi H}\lesssim 1$.

\subsection{SIMP dark matter}

For $m_{h_1}, m_{Z'}\gg m_\chi$ and small couplings between messenger fields and the SM particles, all the $2\rightarrow 2$ annihilation channels are kinematically forbidden or small. Then, the $3\rightarrow 2$ annihilation process for dark matter becomes dominant, namely, $n^2_{\rm DM}\langle\sigma v^2\rangle_{3\rightarrow 2}>n_{\rm DM}\langle\sigma v\rangle_{2\rightarrow 2}$ or $H>n_{\rm DM}\langle\sigma v\rangle_{2\rightarrow 2}$.

The condition for kinetic equilibrium is fulfilled as far as the gauge kinetic mixing is large enough. From $(\sigma v)_{\chi f\rightarrow \chi f}\equiv \frac{\delta^2}{m^2_\chi}$ in eq.~(\ref{kinsection}), it is sufficient to take $|\delta|\simeq 10^{-9}$ for kinetic equilibrium \cite{z3dm}. On the other hand, the SM-annihilating process is subdominant for $n_{\rm DM}\langle\sigma v\rangle_{2\rightarrow 2}<n^2_{\rm DM}\langle\sigma v^2\rangle_{3\rightarrow 2}$ or $n_{\rm DM}\langle\sigma v\rangle_{2\rightarrow 2}<H$, resulting in another condition, $|\delta|\lesssim 10^{-6}$ \cite{z3dm}. 

Consequently, ignoring the $2\rightarrow 2$ annihilation processes, the Boltzmann equation (\ref{genBoltz}) is approximated to
 \bea
\frac{d n_{\rm DM}}{dt} + 3H n_{\rm DM}\approx  -\langle\sigma v^2\rangle_{3\rightarrow 2}(n^3_{\rm DM} -n^2_{\rm DM} n^{\rm eq}_{\rm DM} ). 
\eea

\begin{figure}
  \begin{center}
   \includegraphics[height=0.45\textwidth]{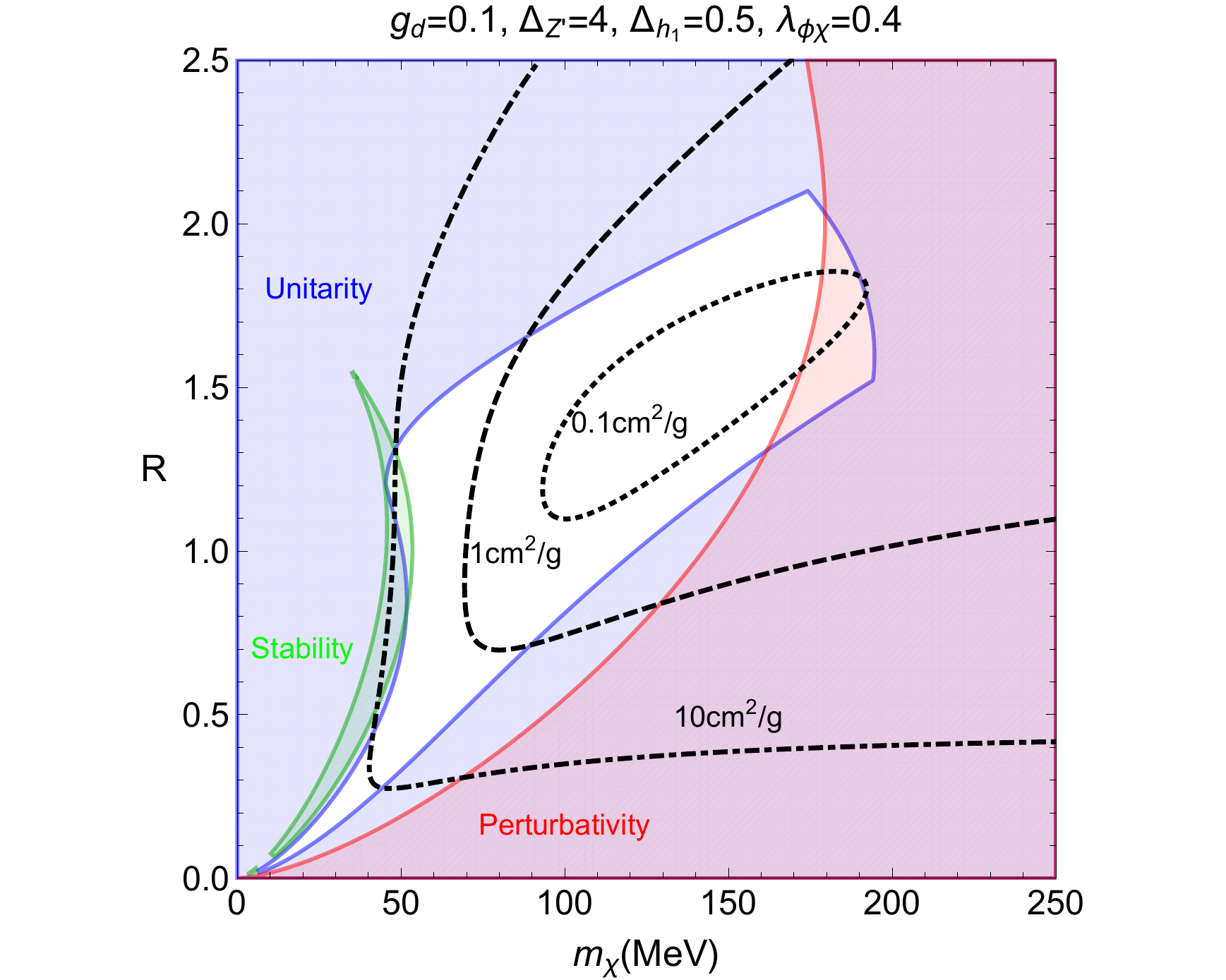}
   \end{center}
  \caption{Parameter space of  $R$(DM cubic coupling) vs $m_\chi$, satisfying the relic density. The regions excluded by unitarity, perturbativity and vacuum stability are shown in blue, red and green, respectively. Dotted, dashed and dot-dashed lines correspond to self-scattering cross sections, $\sigma_{\rm self}/m_\chi=0.1, 1, 10\,{\rm cm^2/g}$. We have chosen $g_d=0.1$, $\lambda_{\phi\chi}=0.4$,  $\Delta_{Z'}=4$ and $\Delta_{h_1}=0.5$.}
  \label{simp-parameter}
\end{figure}

The squared amplitude for $\chi\chi\chi^*\rightarrow \chi^*\chi^*$ scattering is, in the non-relativistic limit, given  \cite{z3dm} by
\bea
|{\cal M}_{\chi\chi\chi^*\rightarrow \chi^*\chi^*}|^2 &=&
\frac{R^2}{16m^2_\chi}\, \bigg( 74\lambda_\chi-117 R^2 -\frac{200 g^2_d m^2_\chi}{ m^2_\chi+m^2_{Z'} } \nonumber \\
&&+\frac{24 \lambda_{\phi\chi} m_\chi^2 (3 m_\chi^2 - 
         2 m_{h_1}^2)-  \lambda_{\phi\chi}^2 (43 m_\chi^2 - 37 m_{h_1}^2) m_{Z'}^2/(9g^2_d)}{(4 m_\chi^2 - m_{h_1}^2) (m_\chi^2 + m_{h_1}^2)} \bigg)^2. 
\eea
Likewise, the squared amplitude for $\chi\chi\chi\rightarrow \chi\chi^*$ scattering is, in the non-relativistic limit, given  \cite{z3dm} by
\bea
|{\cal M}_{\chi\chi\chi\rightarrow \chi\chi^*}|^2 &=&\frac{3R^2}{m^2_\chi}
\bigg(2\lambda_\chi+9R^2 +\frac{25 g^2_d m^2_\chi}{ m^2_\chi+m^2_{Z'} }   \nonumber \\
&&+\frac{2 \lambda_{\phi\chi} m_\chi^2 (13 m_\chi^2 - 
      2 m_{h_1}^2)-   \lambda_{\phi\chi}^2 (19 m_\chi^2 - m_{h_1}^2)m_{Z'}^2/(9g^2_d)} {(9 m_\chi^2 - m_{h_1}^2) (m_\chi^2 + m_{h_1}^2)} \bigg)^2.
\eea

\begin{figure}
  \begin{center}
   \includegraphics[height=0.42\textwidth]{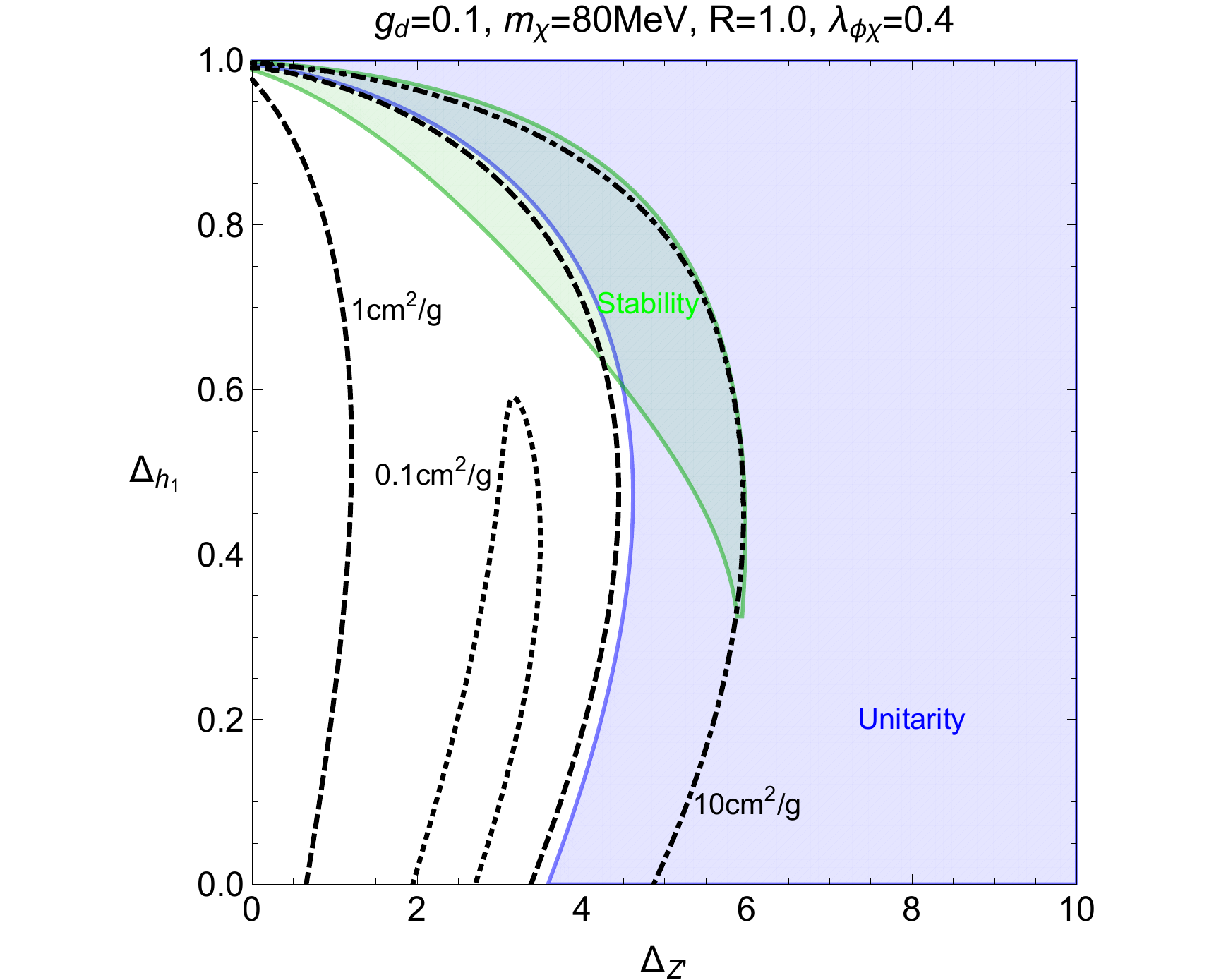} \hspace{-1.2cm}
   \includegraphics[height=0.42\textwidth]{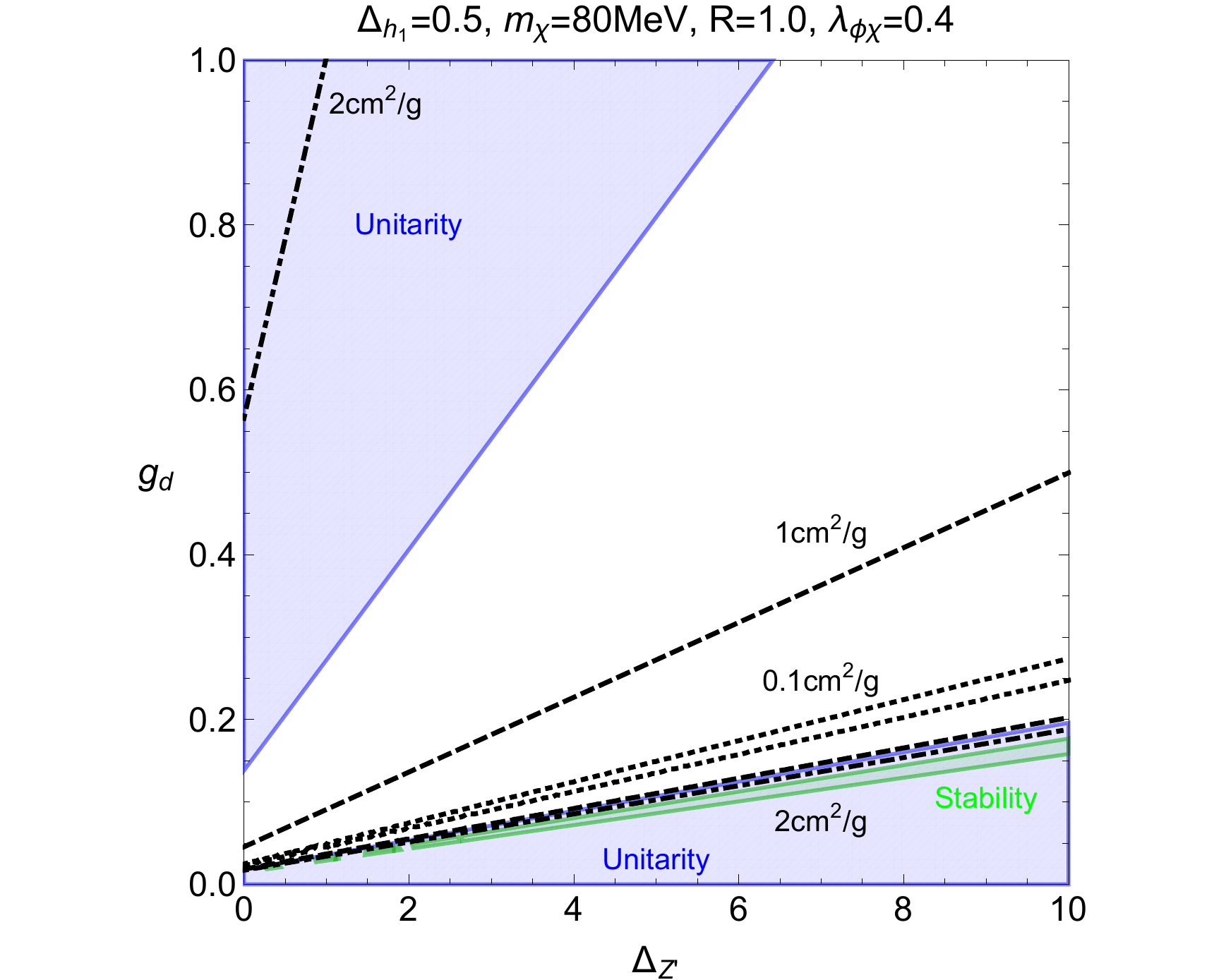}
   \end{center}
  \caption{Parameter space of $\Delta_{Z'}$ vs $\Delta_{h_1}$ (on left) or $\Delta_{Z'}$ vs $g_d$ (on right), satisfying the relic density. The colored regions and  dotted, dashed and dot-dashed lines are as in Fig.~\ref{simp-parameter}, except that the dot-dashed line on right corresponds to $\sigma_{\rm self}/m_\chi=2\,{\rm cm^2/g}$. We have chosen $m_\chi=80\,{\rm MeV}$, $\lambda_{\phi\chi}=0.4$ and $R=1.0$ for all plots and in addition $g_d=0.1$ on left and $\Delta_{h_1}=0.5$ on right. }
  \label{simp-parameter2}
\end{figure}

Then, the effective 3-to-2 annihilation cross section appearing in the above Boltzmann equation is obtained as
 \bea
 \langle\sigma v^2\rangle_{3\rightarrow 2}&=&\frac{1}{4}  \left(\langle \sigma v^2\rangle_{\chi\chi\chi^*\rightarrow \chi^*\chi^*}+ \langle \sigma v^2 \rangle_{\chi\chi\chi\rightarrow \chi\chi^*} \right) \nonumber \\
 &=& \frac{\sqrt{5}}{768\pi m^3_\chi} \bigg( |{\cal M}_{\chi\chi\chi^*\rightarrow \chi^*\chi^*}|^2+ |{\cal M}_{\chi\chi\chi\rightarrow \chi\chi^*}|^2  \bigg) \equiv\frac{\alpha^3_{\rm eff}}{m^5_\chi}. \label{3to2}
 \eea
As a result, solving the Boltzmann equation leads to the DM relic density, given by
\bea
\Omega_{\rm DM} h^2&=&\frac{1.05\times 10^{-10}\,{\rm GeV}^{-2}}{g^{3/4}_* \Big(\frac{m^2_\chi}{M_P}\int^\infty_{x_f}dx\, x^{-5} \langle \sigma v^2\rangle_{3\rightarrow 2} \Big)^{1/2}} \nonumber \\
&=& 1.41\times 10^{-8}\,{\rm GeV}^{-2} \Big(\frac{g_*}{10.75}\Big)^{-3/4}\Big(\frac{x_f}{20}\Big)^2\left(\frac{\alpha_{\rm eff}}{M^{1/3}_P m_\chi}\right)^{-3/2}.
\eea
Therefore, the correct relic density fixes the ratio, $m_\chi/\alpha_{\rm eff}$, which directly predicts the self-scattering cross section, $\sigma_{\rm self}\sim\alpha^2_{\rm eff}/m^2_\chi$.

In Fig.~\ref{simp-parameter}, we have solved the relic density condition for $\lambda_\chi$ and identified the parameter space of $m_\chi$ vs $R$ (DM cubic coupling), that is excluded by unitarity (in blue), perturbativity (in red) and vacuum stability (in green).  Contours with self-scattering cross section with $\sigma_{\rm self}/m_\chi=0.1, 1, 10\,{\rm cm^2/g}$ are shown in black dotted, dashed and dot-dashed lines, respectively.
We have set $g_d=0.1$, $\lambda_{\phi\chi}=0.4$,  $\Delta_{Z'}=4$ and $\Delta_{h_1}=0.5$ where $\Delta_i\equiv (m_i-m_\chi)/m_\chi$.  We find that the newly included vacuum stability bound is less severe than unitarity bound. The dark matter masses are bounded to be smaller than $150\,{\rm MeV}$, due to perturbativity and unitarity. 

On the other hand, in Fig.~\ref{simp-parameter2}, we also drew the parameter space of $\Delta_{Z'}$ vs $\Delta_{h_1}$ on left and $\Delta_{Z'}$ vs $g_d$ on right, that are excluded by unitarity (in blue) and vacuum stability (in green). The black dotted, dashed and dot-dashed lines correspond to contours with self-scattering cross section as in  Fig.~\ref{simp-parameter}, except  that the dot-dashed line on right is for $\sigma_{\rm self}/m_\chi=2\,{\rm cm^2/g}$. As a consequence,  in the allowed parameter space, dark Higgs mass is close to dark matter mass while dark photon mass can be much heavier than dark matter mass. Thus, for relatively light dark Higgs, the forbidden channels, $\chi \chi^*\rightarrow h_1 h_1$ and $\chi\chi\rightarrow h_1 \chi^*$, can be also important in determining the relic density, as will be shown in the later sections. 
From the right plot in  Fig.~\ref{simp-parameter}, the self-scattering cross section can be large, being insensitive to the choice of dark gauge coupling, as far as dark photon mass is large as well.

\section{Thermal freeze-out from forbidden channels}

When dark photon and/or dark Higgs boson masses are close to dark matter mass, they can contribute to the relic density through the forbidden channels, provided that the corresponding $2\rightarrow 2$ cross sections are large enough. Therefore, we still allow for a large self-scattering of dark matter.  We study the parameter space that is consistent with the relic density, first in the case with light dark photon, then the case with light dark Higgs and finally the case where both dark photon and dark Higgs are light. 
Here, we assume that dark photon and/or dark Higgs boson are in kinetic equilibrium during the freeze-out process.

\subsection{The case with $m_\chi<m_{Z'}\ll m_{h_1}$}

\begin{figure}
  \begin{center}
   \vspace{-2.5cm}
   \includegraphics[height=0.35\textwidth]{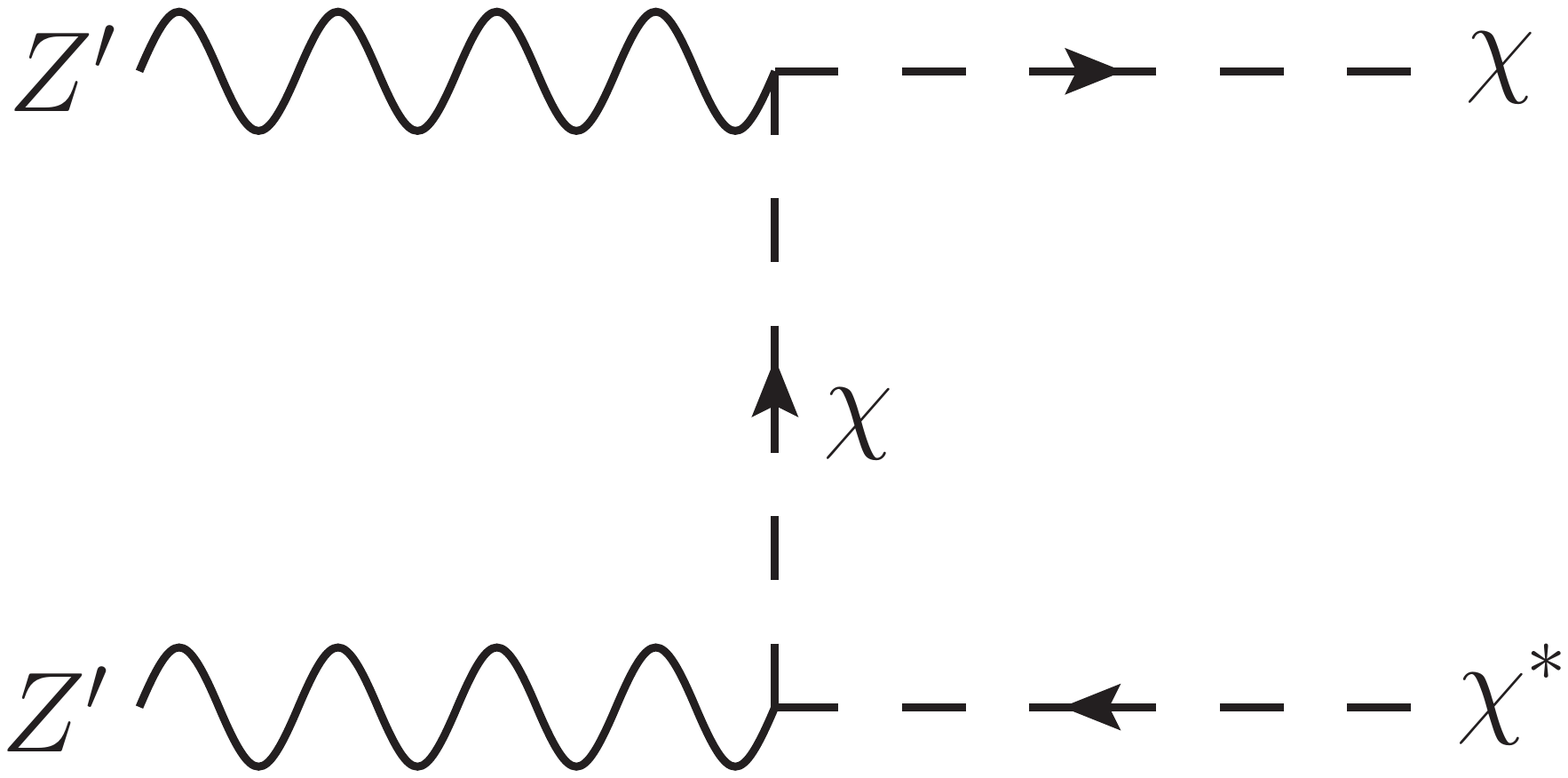}
  \includegraphics[height=0.35\textwidth]{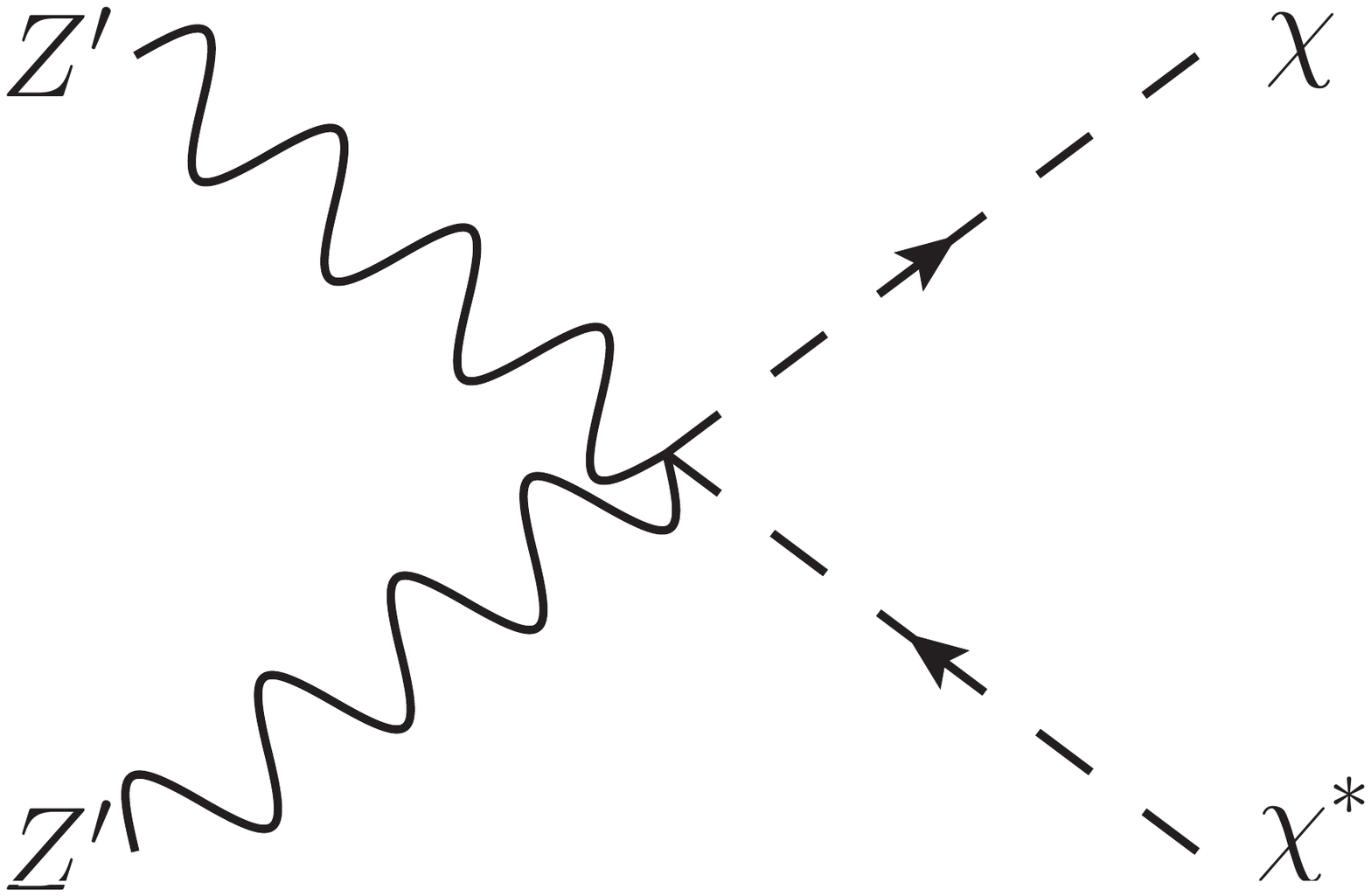}
   \includegraphics[height=0.35\textwidth]{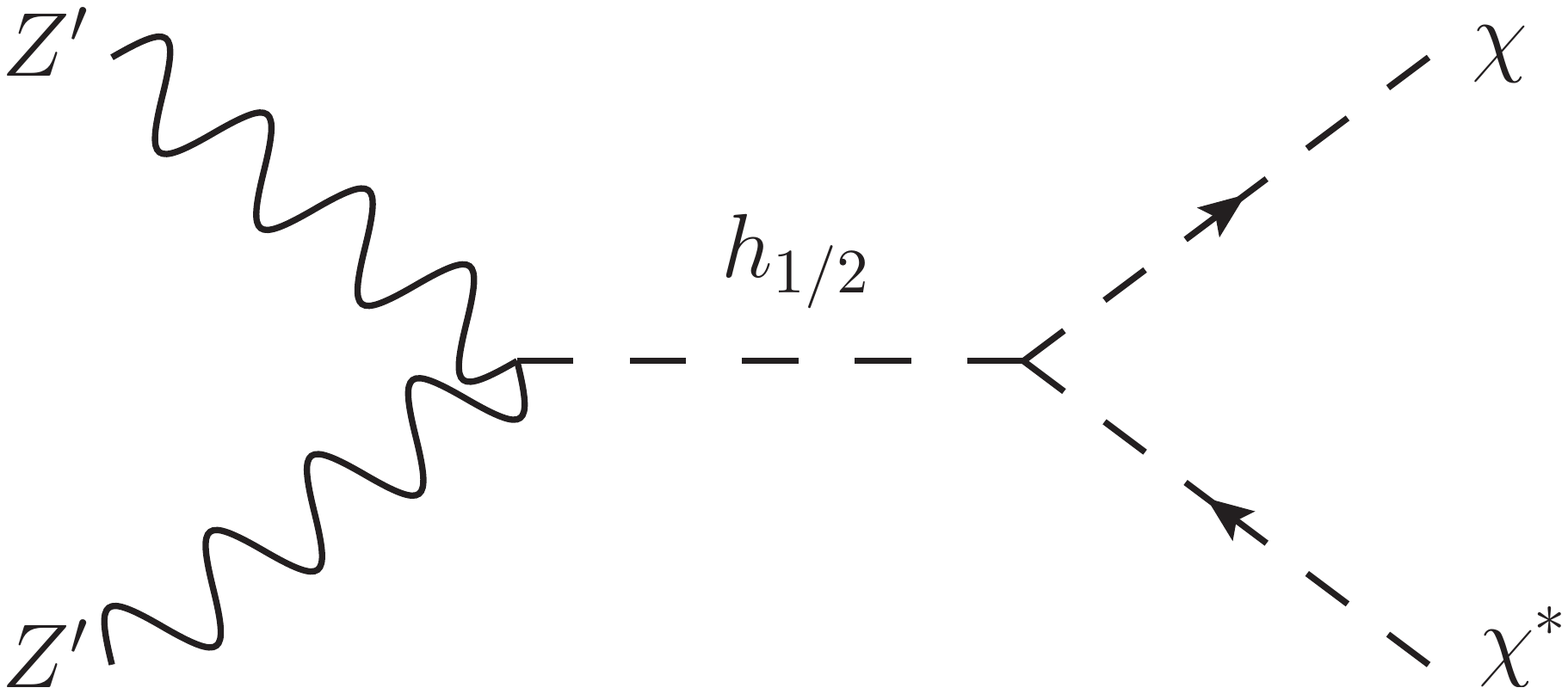}  \vspace{-2.5cm} \\
   \includegraphics[height=0.35\textwidth]{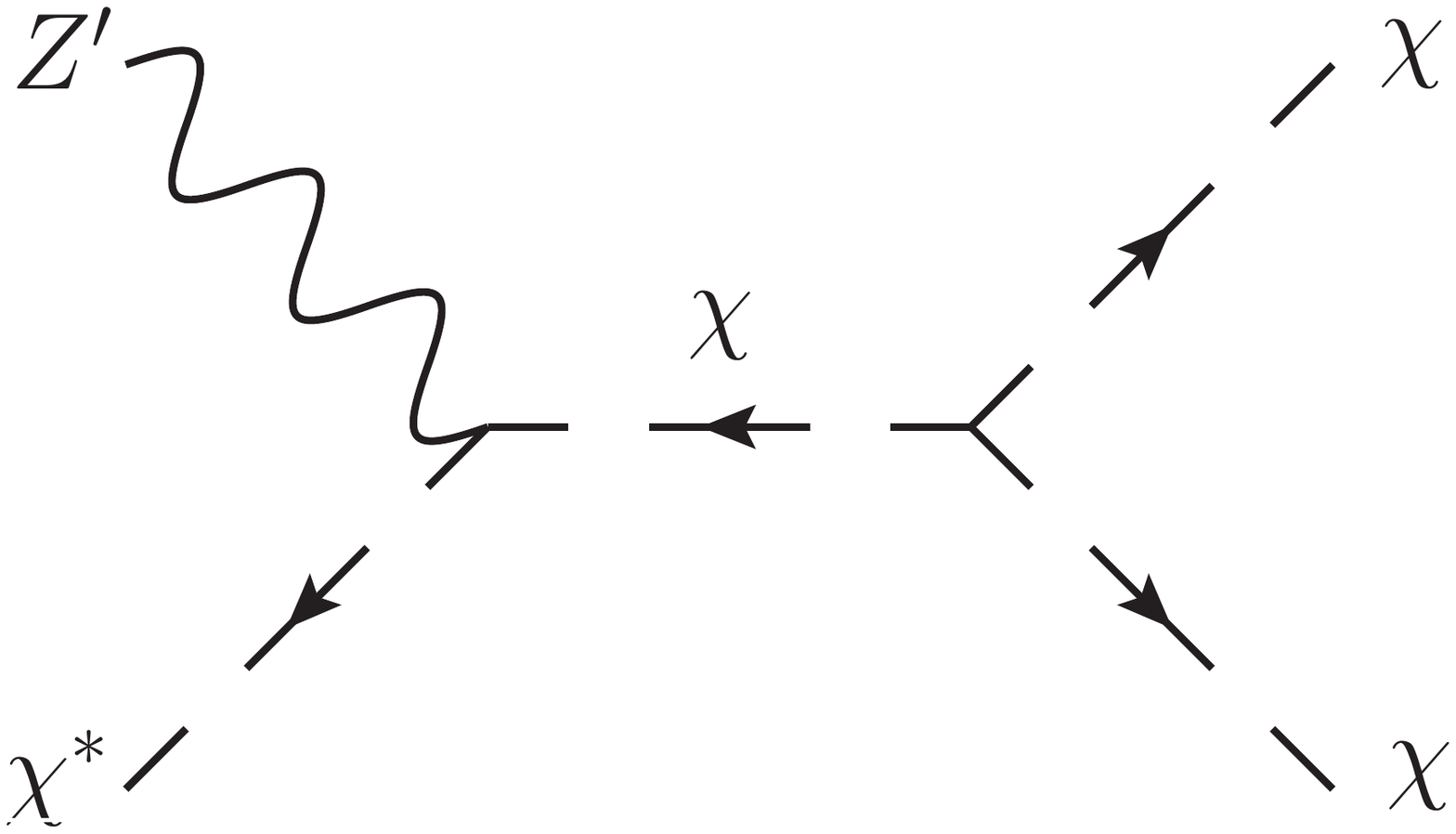}
    \includegraphics[height=0.35\textwidth]{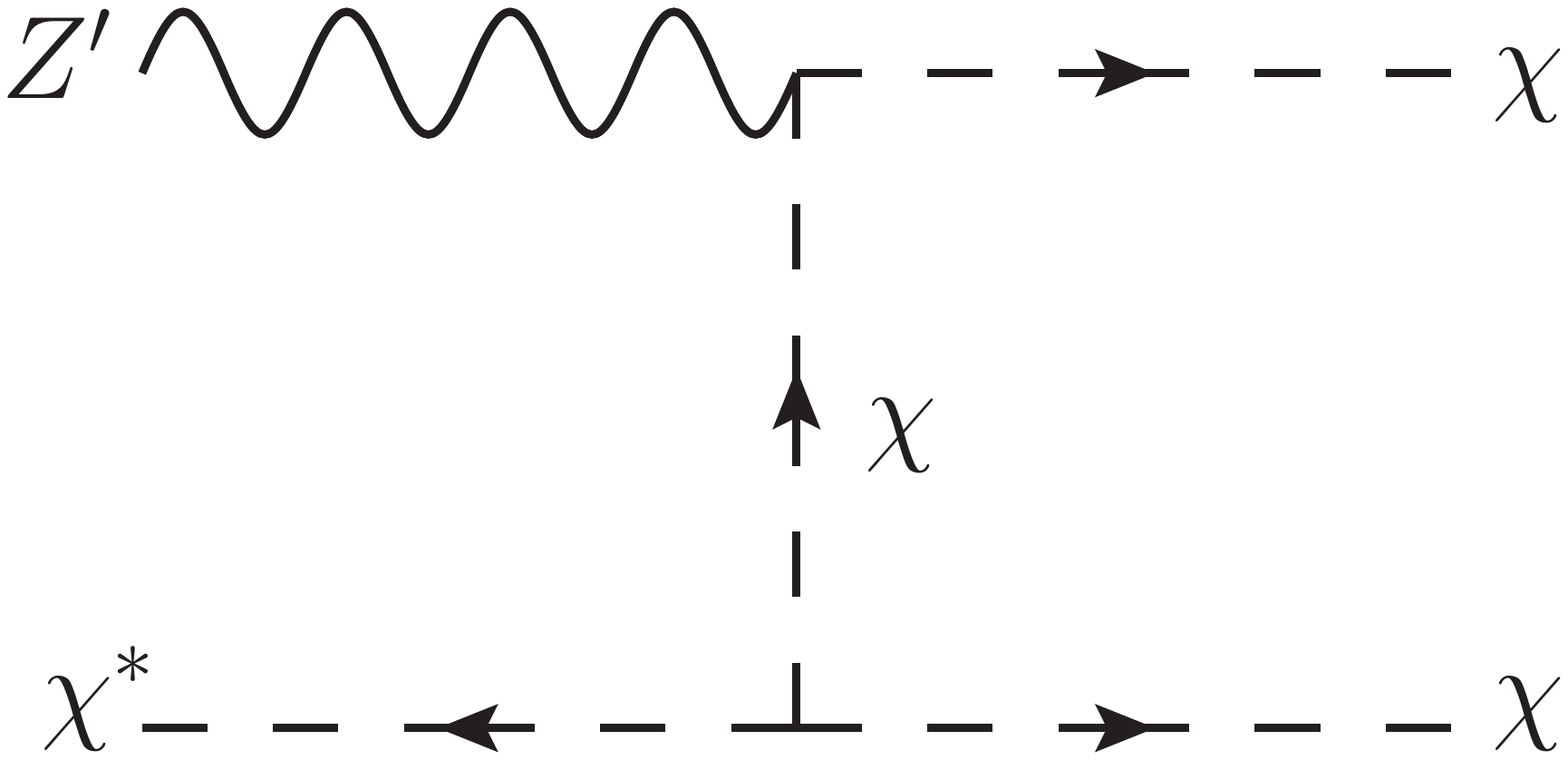}   \vspace{-1.5cm}
   \end{center}
  \caption{Feynmann diagrams for forbidden channels with $Z'$.}
  \label{Zp-forb}
\end{figure}

In the case where $m_\chi<m_{Z'}$ and dark Higgs is much heavier than the other particles, the forbidden channels involving $Z'$ as shown in Fig.~\ref{Zp-forb} contribute to the Boltzmann equation.
Then, the Boltzmann equation (\ref{genBoltz}) is approximated to 
 \bea
\frac{d n_{\rm DM}}{dt} + 3H n_{\rm DM}&\approx&-\frac{1}{2}\langle \sigma v\rangle_{\chi\chi^*\rightarrow Z'Z'} n^2_{\rm DM}+2\langle\sigma v\rangle_{Z'Z'\rightarrow \chi\chi^*} (n^{\rm eq}_{Z'})^2 \nonumber \\
&& -\frac{1}{2}\langle\sigma v\rangle_{\chi\chi\rightarrow Z'\chi^*} n^2_{\rm DM}+\langle \sigma v\rangle_{Z'\chi^*\rightarrow \chi\chi} n^{\rm eq}_{Z'} n_{\rm DM}.  
 \eea

The detailed balance conditions at high temperature are
\bea
\langle\sigma v\rangle_{\chi\chi^*\rightarrow Z'Z'} &=&\frac{4 (n^{\rm eq}_{Z'})^2}{(n^{\rm eq}_{\rm DM})^2}\langle\sigma v\rangle_{Z'Z'\rightarrow \chi\chi^*} \nonumber \\
&=& 9 (1+\Delta_{Z'})^3 e^{-2\Delta_{Z'} x}\, \langle\sigma v\rangle_{Z'Z'\rightarrow \chi\chi^*} \label{Zp1}
\eea
and 
\bea
\langle\sigma v\rangle_{\chi\chi\rightarrow Z'\chi^*} &=&\frac{2n^{\rm eq}_{Z'}}{n^{\rm eq}_{\rm DM}} \langle \sigma v\rangle_{Z'\chi^*\rightarrow \chi\chi}  \nonumber \\
&=&3 (1+\Delta_{Z'})^{3/2} e^{-\Delta_{Z'} x}\,\langle \sigma v\rangle_{Z'\chi^*\rightarrow \chi\chi}  \label{Zp2}
\eea
with $\Delta_{Z'}\equiv (m_{Z'}-m_\chi)/m_\chi$.

\begin{figure}
  \begin{center}
   \includegraphics[height=0.42\textwidth]{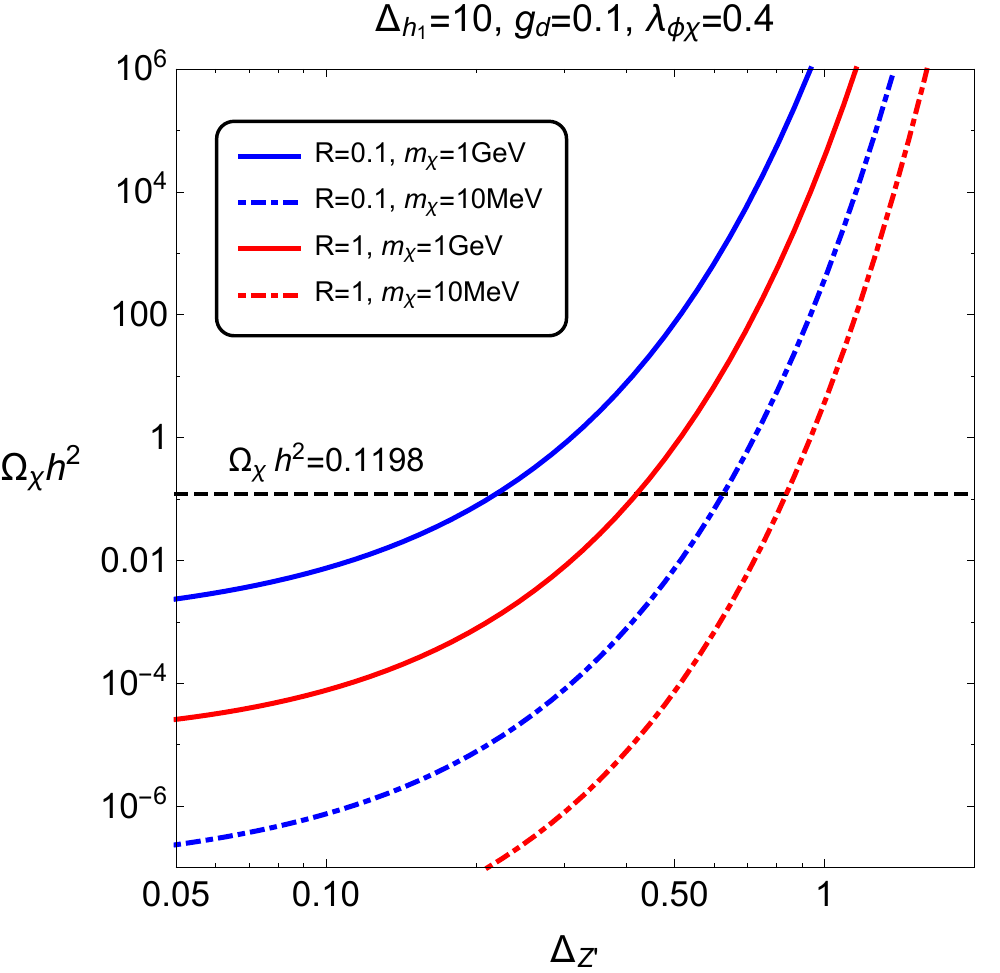}
      \includegraphics[height=0.42\textwidth]{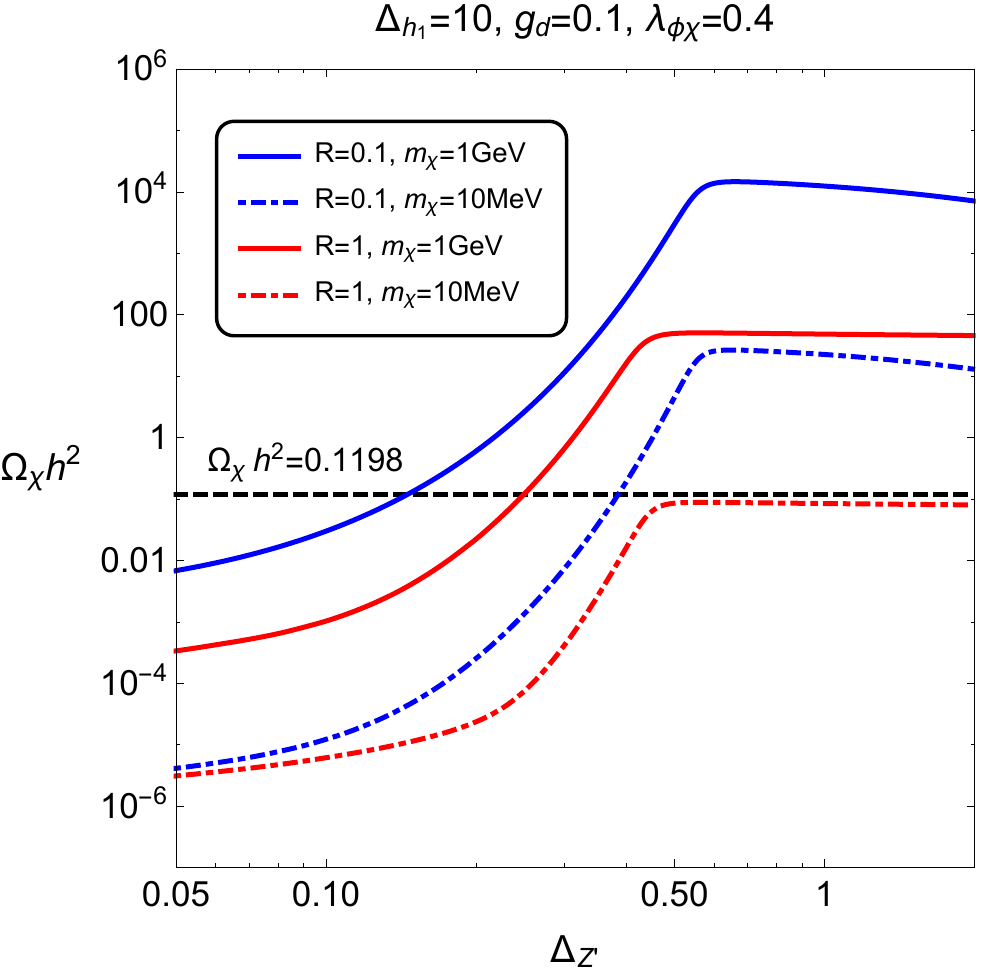}
   \end{center}
  \caption{Dark matter relic density as a function of  $\Delta_{Z'}$ for only forbidden channels with $Z'$ on left and both SIMP and forbidden channels with $Z'$ on right. 
 We have taken $(R,m_\chi)=(0.1, m_\chi=1\,{\rm GeV})$, $(0.1, m_\chi=10\,{\rm MeV})$, $(1.0, m_\chi=1\,{\rm GeV})$ and $(1.0, m_\chi=10\,{\rm MeV})$, from top to bottom. Black dashed lines correspond to the central value of relic density, $\Omega_\chi h^2=0.1198$, from Planck.  In both plots, we chose $g_d=0.1$, $\lambda_{\phi\chi}=0.4$ and $\Delta_{h_1}=10$. }
  \label{forbZ}
\end{figure}

Then, we can rewrite the Boltzmann equation by using the detailed balance conditions, (\ref{Zp1}) and (\ref{Zp2}),  as follows,
\bea
\frac{dY_{\rm DM}}{dx}&=&-\lambda x^{-2}  \langle\sigma v\rangle_{Z'Z'\rightarrow \chi\chi^*} \left(\frac{9}{2}(1+\Delta_{Z'})^3 e^{-2\Delta_{Z'}x} \,Y^2_{\rm DM}-2(Y^{\rm eq}_{Z'})^2
\right) \nonumber \\
&&-\lambda x^{-2}  \langle\sigma v\rangle_{Z'\chi^*\rightarrow \chi\chi}  \left( \frac{3}{2}(1+\Delta_{Z'})^{3/2} e^{-\Delta_{Z'} x}\, Y^2_{\rm DM}- Y^{\rm eq}_{Z'} Y_{\rm DM}\right) \label{Zp-Boltz}
\eea
where $\lambda\equiv s(m_\chi)/H(m_\chi)$ with $s(m_\chi)=\frac{2\pi^2}{45}\,g_{*s} m^3_\chi$ and $1/H(m_\chi)=3.02 g^{-1/2}_* \frac{M_P}{m^2_\chi}$. 
As a result, the approximate solution to the Boltzmann equation (\ref{Zp-Boltz}) is given by
\bea
(Y_{\rm DM}(\infty))^{-1}
&=&\lambda \int^\infty_{x_f} dx \, x^{-2}\Big(\frac{9}{2}(1+\Delta_{Z'})^3 e^{-2\Delta_{Z'}x} \langle\sigma v\rangle_{Z'Z'\rightarrow \chi\chi^*} \nonumber \\
&& +   \frac{3}{2}(1+\Delta_{Z'})^{3/2} e^{-\Delta_{Z'} x}   \langle\sigma v\rangle_{Z'\chi^*\rightarrow \chi\chi}  \Big)
\eea

\begin{figure}
  \begin{center}
      \includegraphics[height=0.45\textwidth]{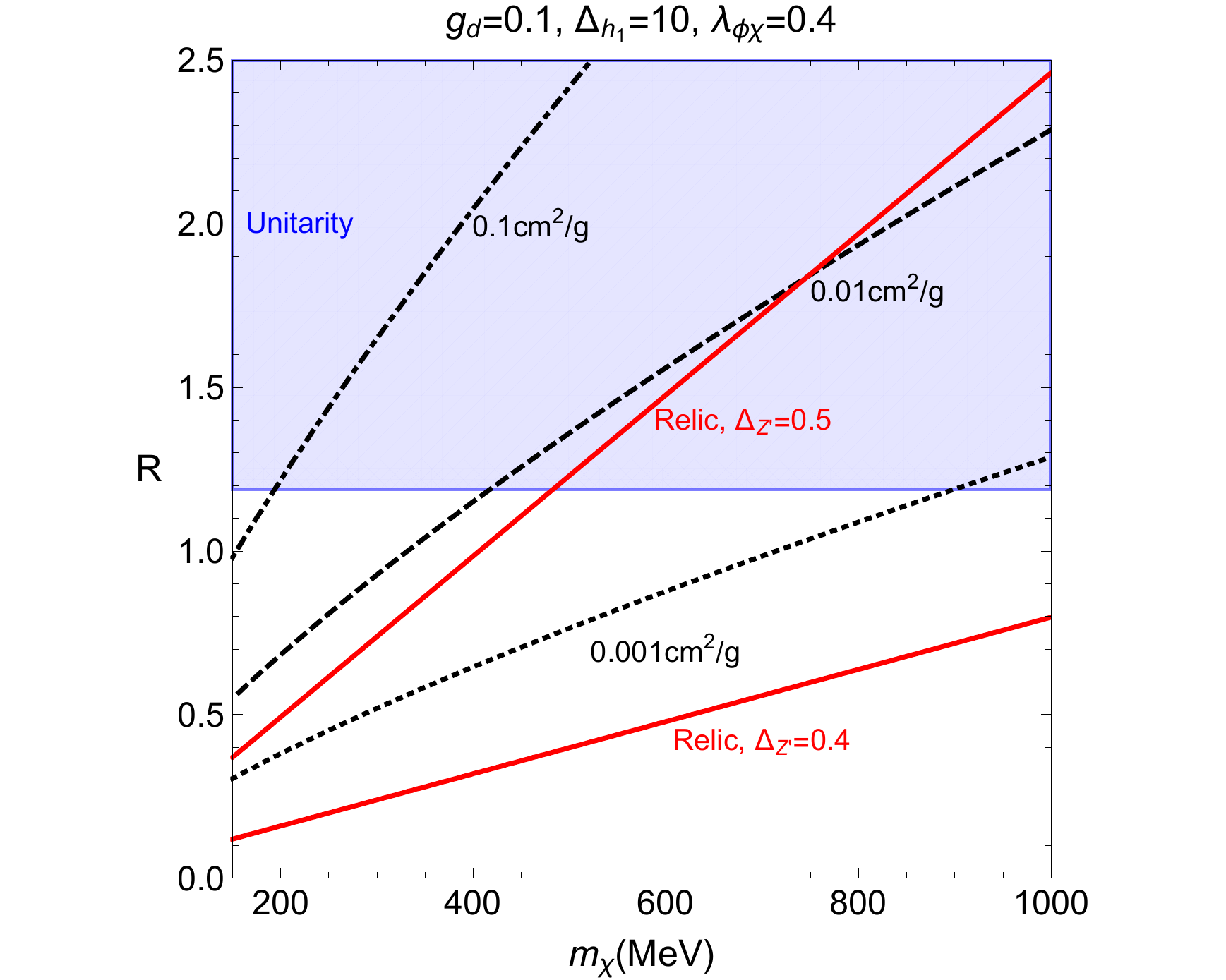}
   \end{center}
  \caption{Parameter space of $R$ vs $m_\chi$ for forbidden channels with $Z'$. The red lines satisfy the relic density and the blue region is excluded by unitarity. Dotted, dashed and dot-dashed lines correspond to self-scattering cross sections, $\sigma_{\rm self}/m_\chi=0.001, 0.01, 0.1\,{\rm cm^2/g}$. We took $\Delta_{Z'}=0.4$ or $ 0.5$ and the value of $\lambda_\chi$ saturates the vacuum stability bound, and $g_d=0.1$, $\lambda_{\phi\chi}=0.4$ and $\Delta_{h_1}=10$. }
  \label{forbZ2}
\end{figure}

Setting $(\sigma v)_{Z'Z'\rightarrow \chi\chi^*} =a$ and $ (\sigma v)_{Z'\chi^*\rightarrow \chi\chi}=b v^2 $, which leads to $\langle\sigma v\rangle_{Z'Z'\rightarrow \chi\chi^*} =a$ and $ \langle\sigma v\rangle_{Z'\chi^*\rightarrow \chi\chi}=6b/x$, where the detailed expressions for $a$ and $b$ are given in eqs.~(\ref{zpzp}) and (\ref{zpchi}),
we get the DM abundance as
\be
Y_{\rm DM}(\infty)\approx \frac{x_f}{\lambda}\, e^{\Delta_{Z'} x_f}\, g(\Delta_{Z'}, x_f)
\ee
with
\bea
g(\Delta_{Z'},x_f)&=& \bigg[\frac{9b}{2x_f}\,(1+\Delta_{Z'})^{3/2}\Big(1-(\Delta_{Z'}x_f)^2\, e^{\Delta_{Z'}x_f} \int^\infty_{\Delta_{Z'}x_f} dt \,t^{-2} e^{-t}\Big) \nonumber \\
&&+\frac{9a}{2}\,(1+\Delta_{Z'})^3  e^{-\Delta_{Z'} x_f} \Big(1-2(\Delta_{Z'}x_f)\, e^{2\Delta_{Z'}x_f}  \int^\infty_{2\Delta_{Z'}x_f} dt \,t^{-1} e^{-t} \Big) \bigg]^{-1}. 
\eea
Consequently, the relic density is determined to be
\bea
\Omega_{\rm DM} h^2 
=5.20\times 10^{-10} {\rm GeV}^{-2} \Big(\frac{g_*}{10.75} \Big)^{-1/2}\Big(\frac{x_f}{20}\Big)\, e^{\Delta_{Z'} x_f}\, g(\Delta_{Z'}, x_f).
\eea
Then, the $2\rightarrow 2$ annihiation cross sections can be large, due to the inverse of the Boltzmann suppression factor, $e^{\Delta_{Z'} x_f}$, appearing in the relic density. 
Therefore, the self-scattering cross section of dark matter can be large enough. But, for a small $\Delta_{Z'}$, dark matter self-interaction can be smaller than in the SIMP case, being compatible with the relic density.

In Fig.~\ref{forbZ}, we depicted the relic density as a function of $\Delta_{Z'}$, only with forbidden channels involving $Z'$ on left and with both SIMP and forbidden channels on right, varying the self-interaction and mass of dark matter, $(R, m_\chi)$, between $R=0.1-1$ and $m_\chi=10\,{\rm MeV}-1\,{\rm GeV}$.  For both plots, we took $g_d=0.1$, $\lambda_{\phi\chi}=0.4$ and $\Delta_{h_1}=10$. 
On the right plot of Fig.~\ref{forbZ}, as $\Delta_{Z'}$ gets larger than about $0.5$ the relic density approaches a certain fixed value, that is determined by the SIMP processes dominantly.   However, for $\Delta_{Z'}\lesssim 0.5$, the relic density become sensitive to the value of $\Delta_{Z'}$, as well as to $R$ and $m_\chi$.

In Fig.~\ref{forbZ2}, we also show the parameter space of $m_\chi$ vs $R$, satisfying the relic density in red lines, for $\Delta_{Z'}=0.4$ and $0.5$, from bottom to top. We took dark matter masses to be larger than $150\,{\rm MeV}$ to cover beyond the maximal value allowed by unitarity in the SIMP case. Furthermore, we chose the value of $\lambda_\chi$ such that the vacuum stability bound is saturated and set $g_d=0.1$, $\lambda_{\phi\chi}=0.4$ and $\Delta_{h_1}=10$.  The blue region is excluded by unitarity and the contours with self-scattering cross section, $\sigma_{\rm self}/m_\chi=0.001, 0.01, 0.1\,{\rm cm^2/g}$, are shown in dotted, dashed and dot-dashed lines, respectively. For relatively heavy DM masses, the self-scattering cross section is smaller than in the SIMP case, being consistent with the relic density and unitarity. Therefore, the forbidden channels are crucial to keep the model perturbative for the wide range of masses for light dark matter.

\subsection{The case with $m_\chi<m_{h_1}\ll m_{Z'}$}

\begin{figure}
  \begin{center}
   \vspace{-2.5cm}
   \includegraphics[height=0.35\textwidth]{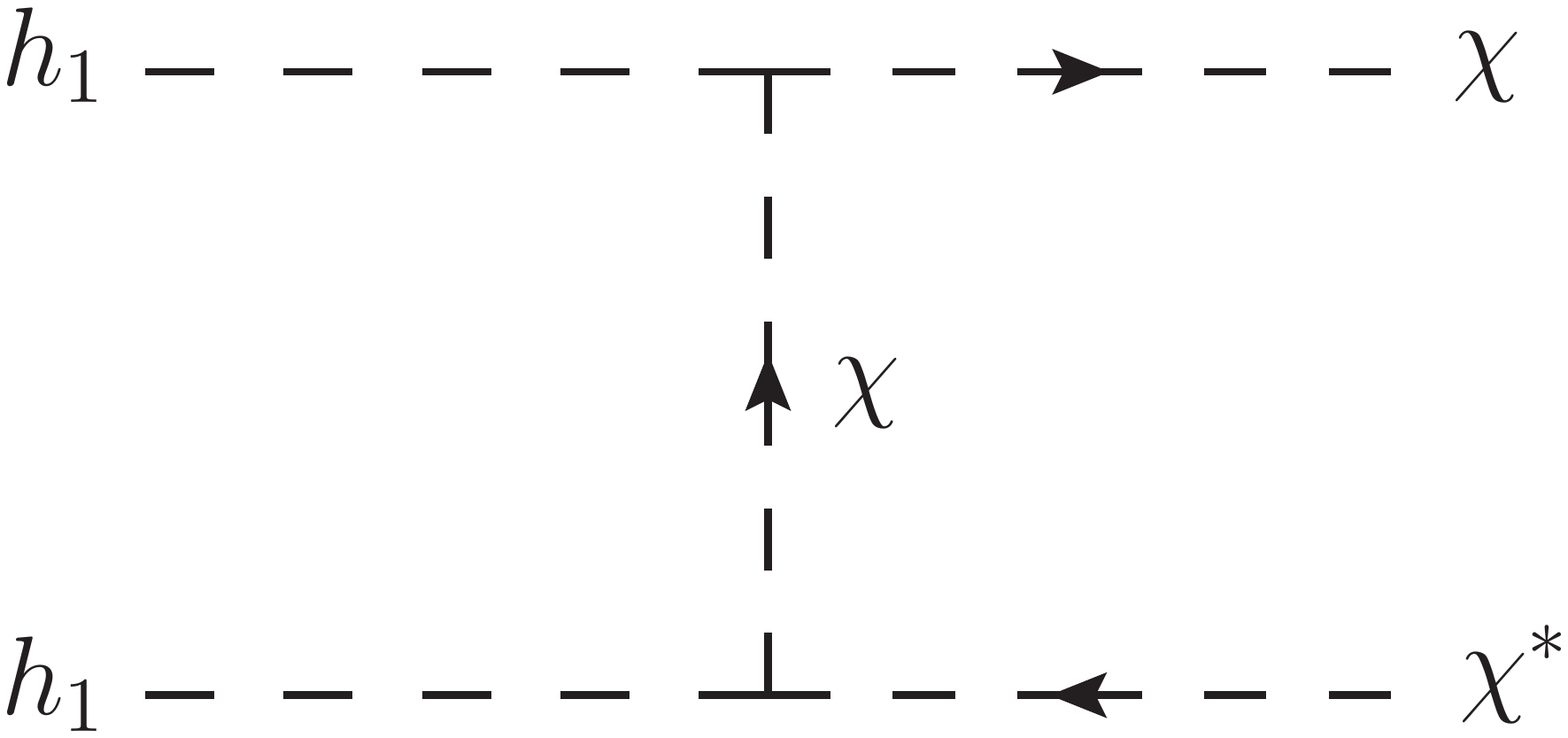}
    \includegraphics[height=0.35\textwidth]{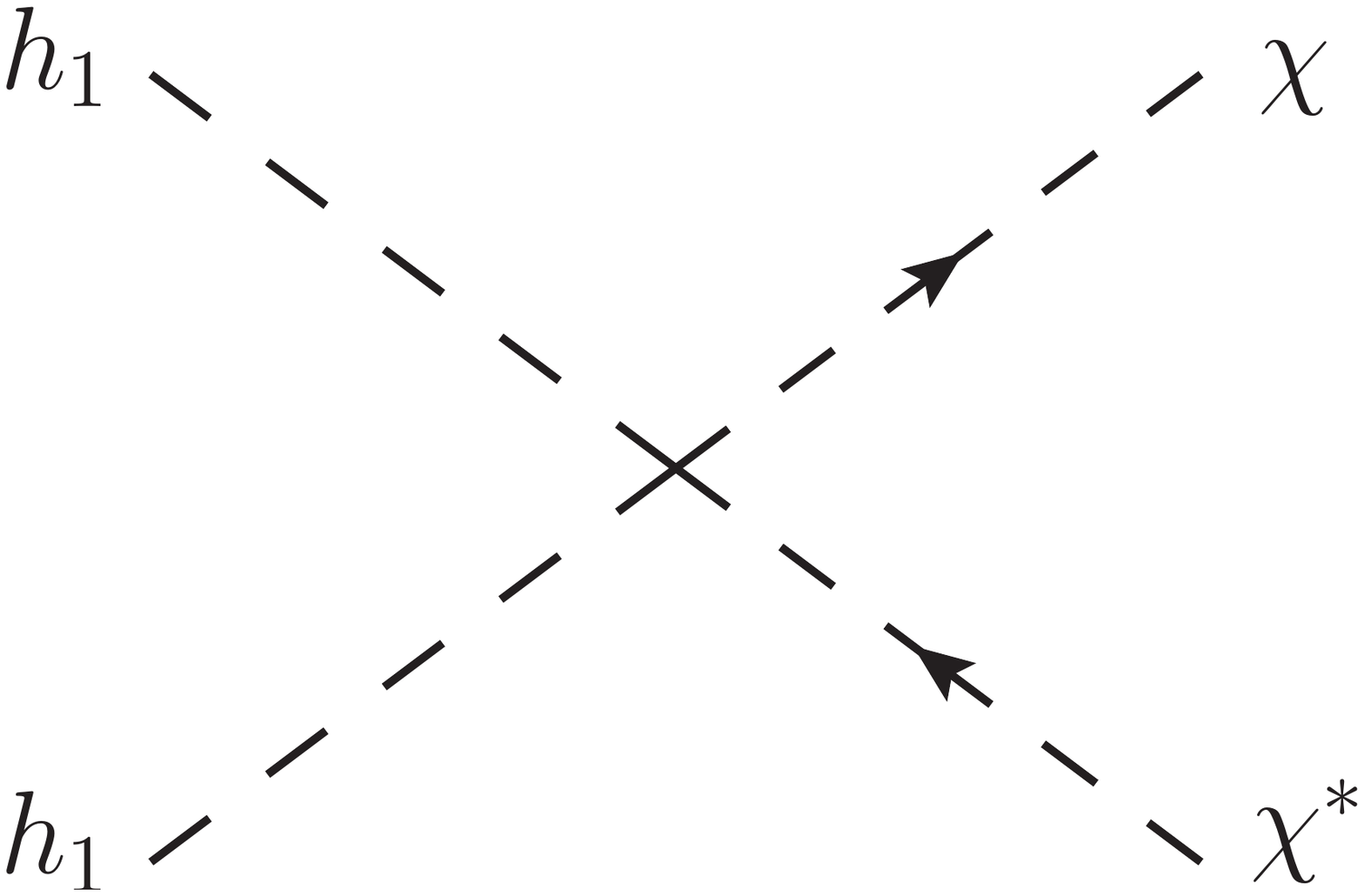} 
    \includegraphics[height=0.35\textwidth]{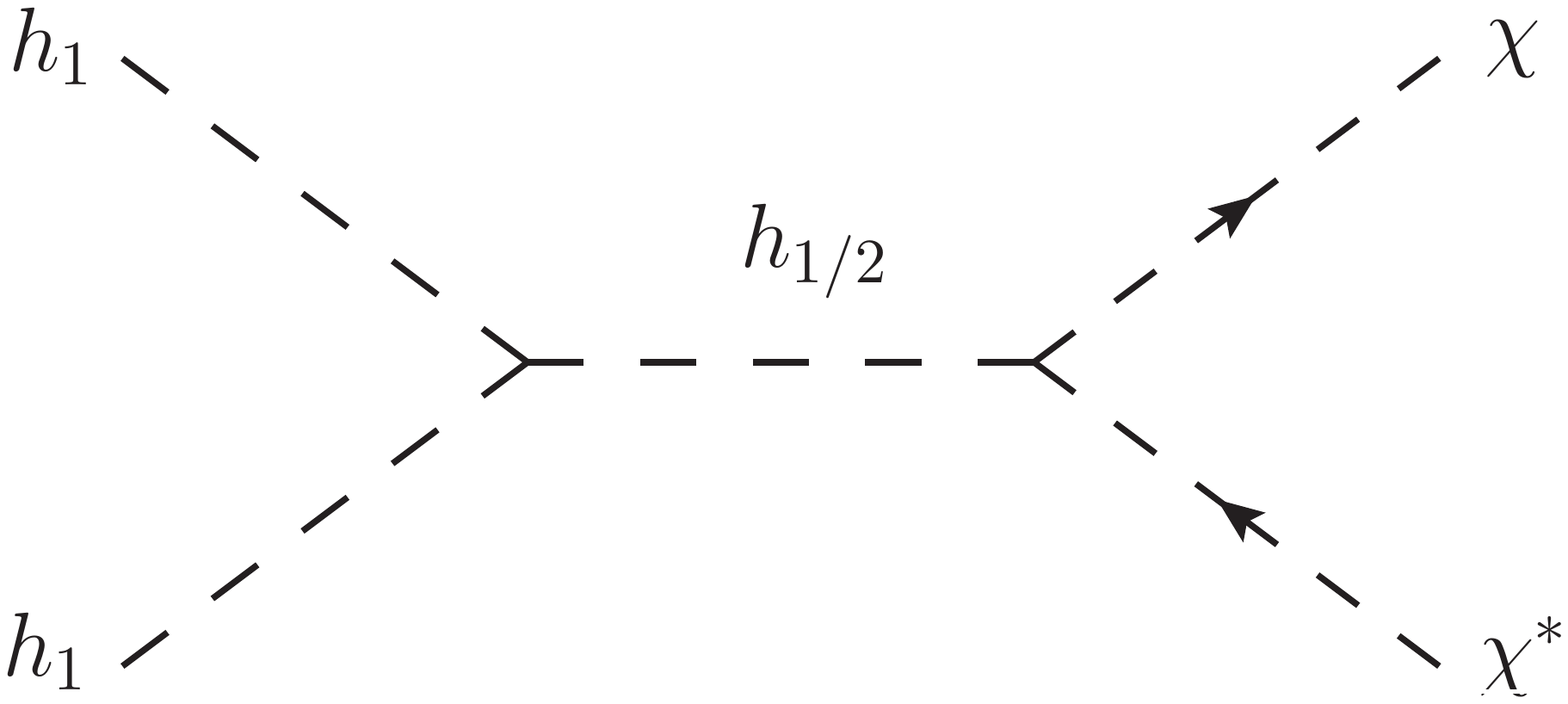}  \vspace{-2.5cm}  \\
 \includegraphics[height=0.35\textwidth]{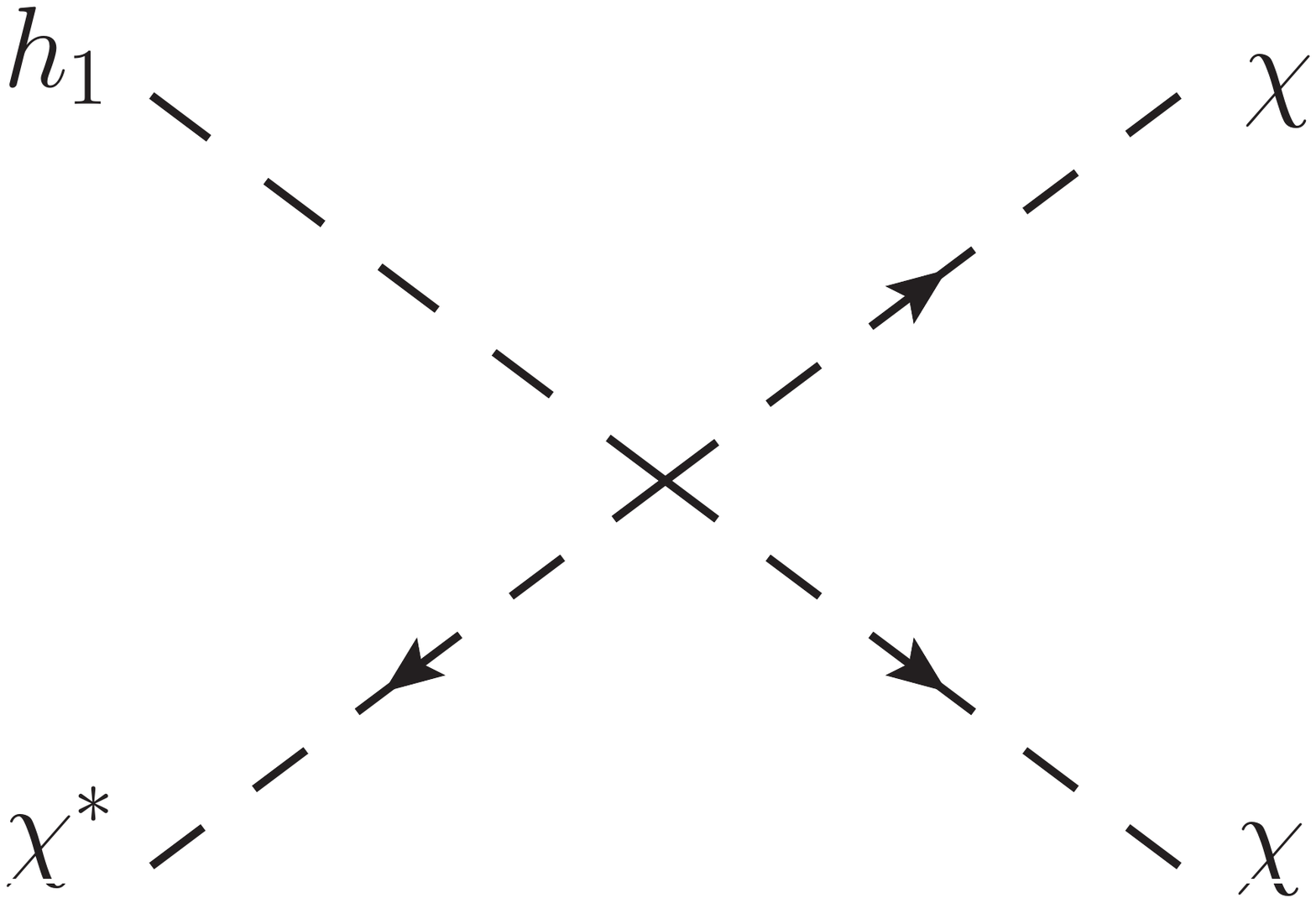}
  \includegraphics[height=0.35\textwidth]{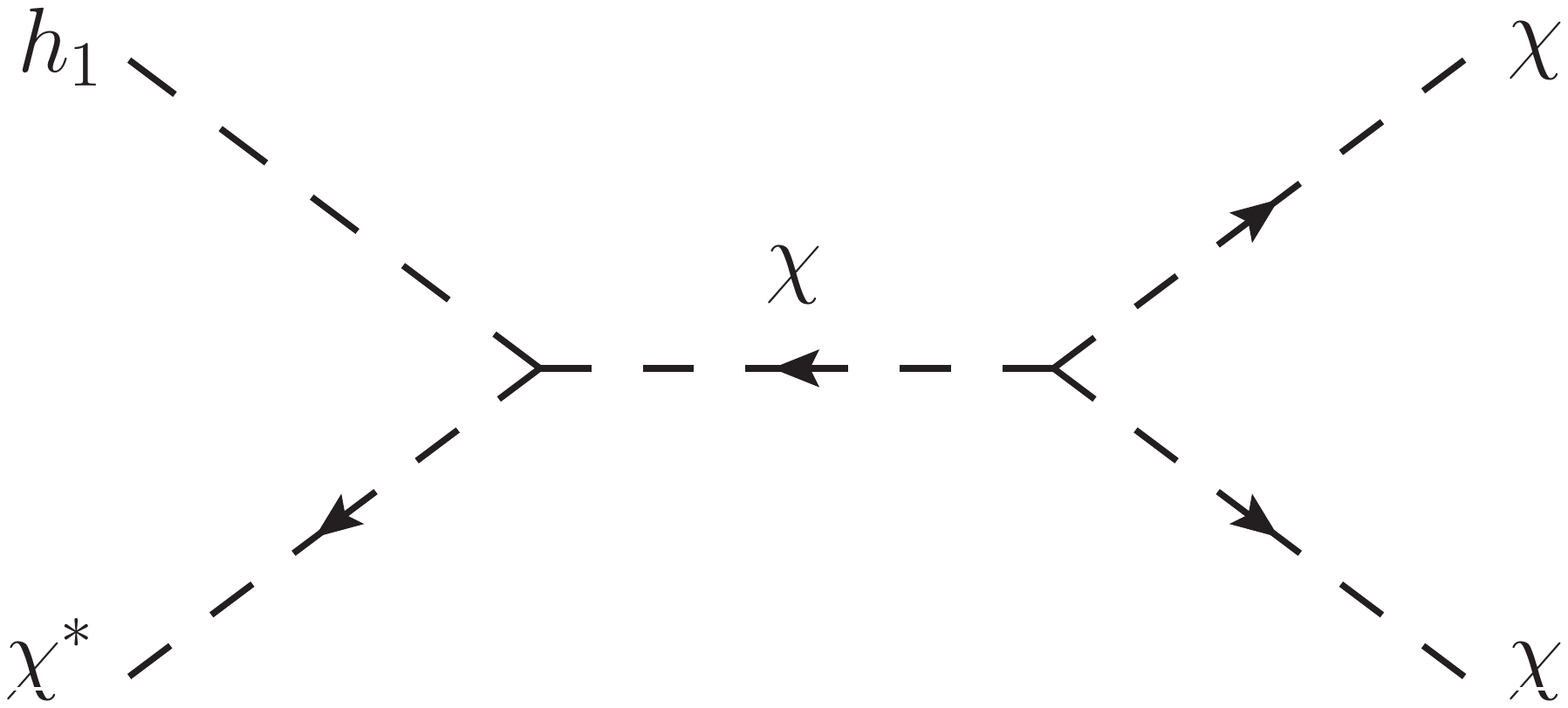}
   \includegraphics[height=0.35\textwidth]{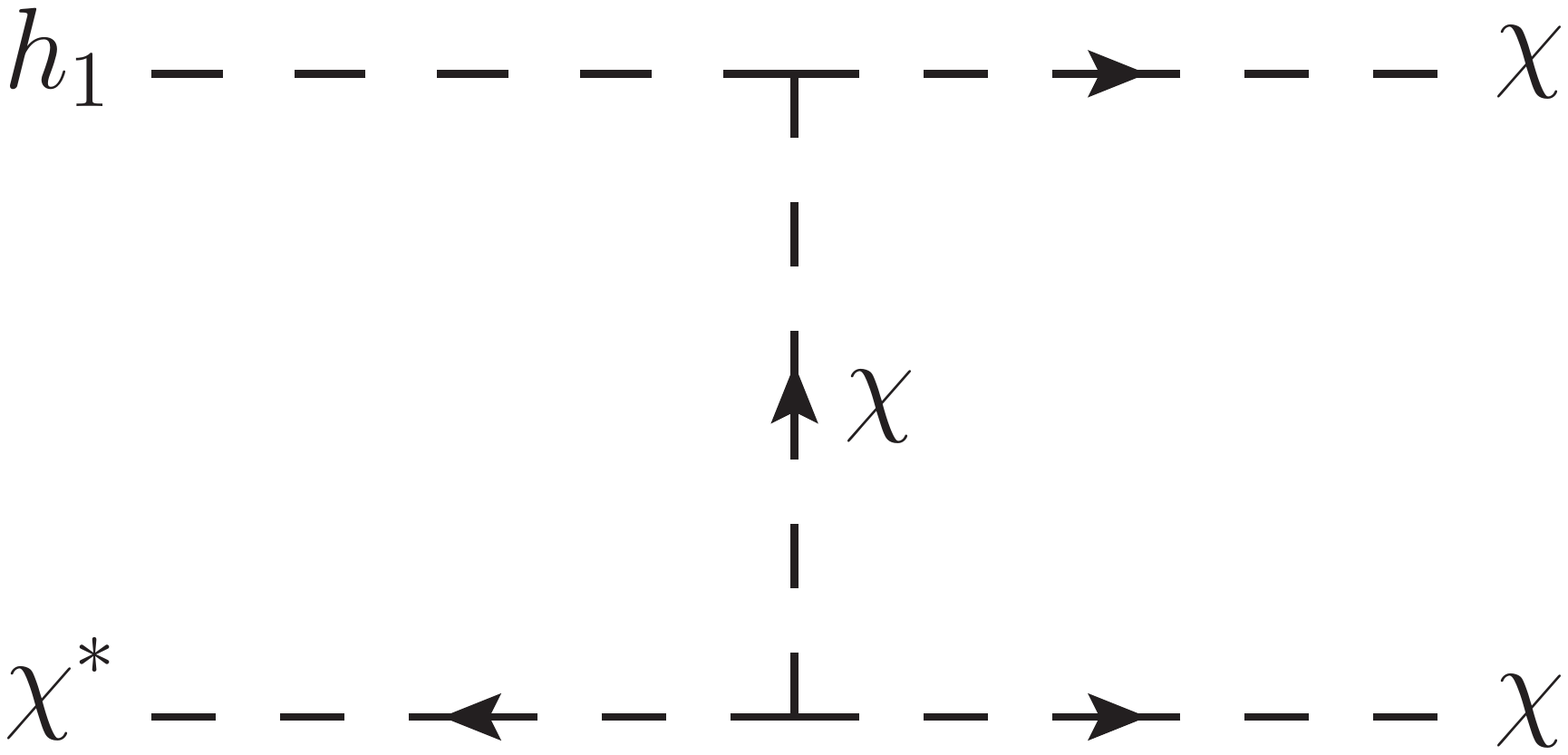}  \vspace{-1.5cm} 
   \end{center}
  \caption{Feynmann diagrams for forbidden channels with $h_1$.}
  \label{h1-forb}
\end{figure}

In the case where $m_\chi<m_{h_1}$ and dark photon is much heavier than the other particles,  the forbidden channels involving $h_1$ as shown in Fig.~\ref{h1-forb} contribute to determining the relic density.  In this case, the Boltzmann equation (\ref{genBoltz}) is approximated to 
 \bea
\frac{d n_{\rm DM}}{dt} + 3H n_{\rm DM}
&\approx&-\frac{1}{2}\langle \sigma v\rangle_{\chi\chi^*\rightarrow h_1 h_1} n^2_{\rm DM}+2\langle\sigma v\rangle_{h_1h_1\rightarrow \chi\chi^*} (n^{\rm eq}_{h_1})^2 \nonumber \\
&& -\frac{1}{2}\langle\sigma v\rangle_{\chi\chi\rightarrow h_1\chi^*} n^2_{\rm DM}+\langle \sigma v\rangle_{h_1\chi^*\rightarrow \chi\chi} n^{\rm eq}_{h_1} n_{\rm DM}. 
 \eea
Similarly to the case with light $Z'$,  the detailed balance conditions at high temperature are
\bea
\langle\sigma v\rangle_{\chi\chi^*\rightarrow h_1 h_1} &=&\frac{4 (n^{\rm eq}_{h_1})^2}{(n^{\rm eq}_{\rm DM})^2}\langle\sigma v\rangle_{h_1h_1\rightarrow \chi\chi^*} \nonumber \\
&=&  (1+\Delta_{h_1})^3 e^{-2\Delta_{h_1} x}\, \langle\sigma v\rangle_{h_1 h_1\rightarrow \chi\chi^*}\label{H1H1}
\eea
and 
\bea
\langle\sigma v\rangle_{\chi\chi\rightarrow h_1\chi^*} &=&\frac{2n^{\rm eq}_{h_1}}{n^{\rm eq}_{\rm DM}} \langle \sigma v\rangle_{h_1\chi^*\rightarrow \chi\chi}  \nonumber \\
&=&(1+\Delta_{h_1})^{3/2} e^{-\Delta_{h_1} x}\,\langle \sigma v\rangle_{h_1\chi^*\rightarrow \chi\chi}\label{H1chi} 
\eea
with $\Delta_{h_1}\equiv (m_{h_1}-m_\chi)/m_\chi$.

\begin{figure}
  \begin{center}
   \includegraphics[height=0.42\textwidth]{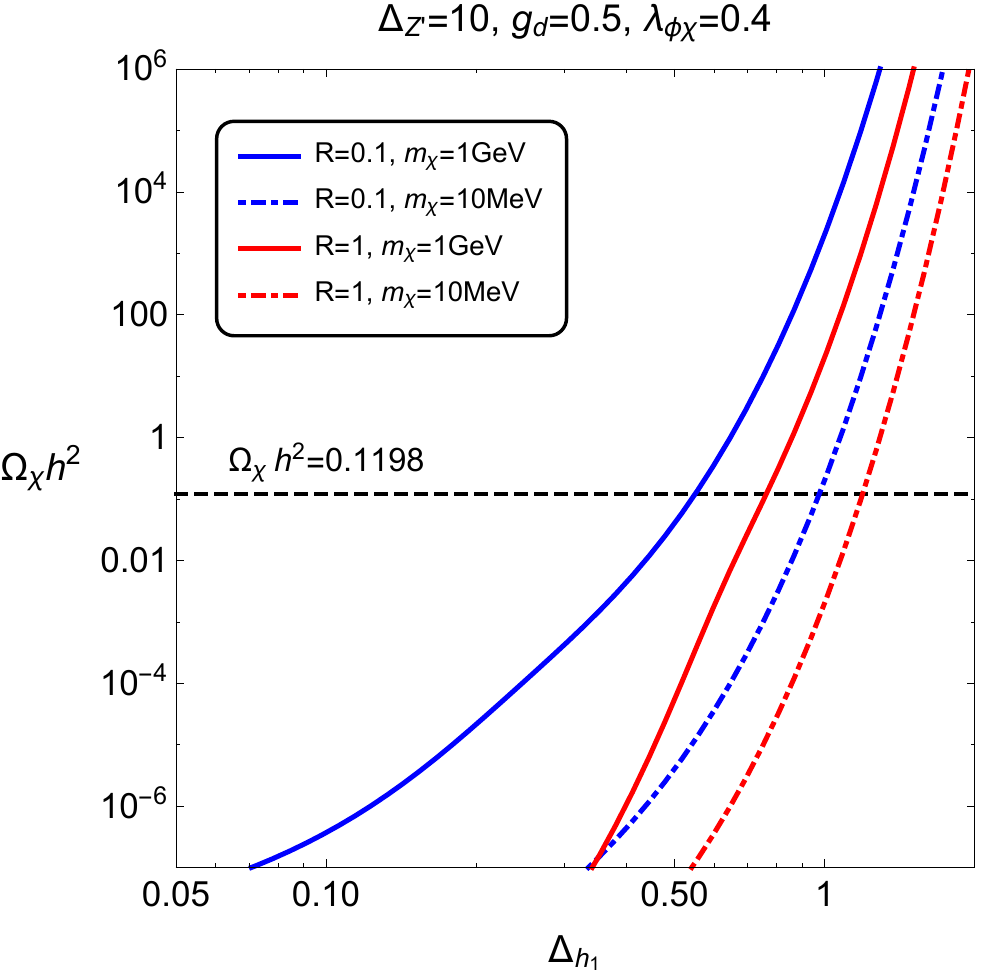} ~~~~~~~~~
      \includegraphics[height=0.42\textwidth]{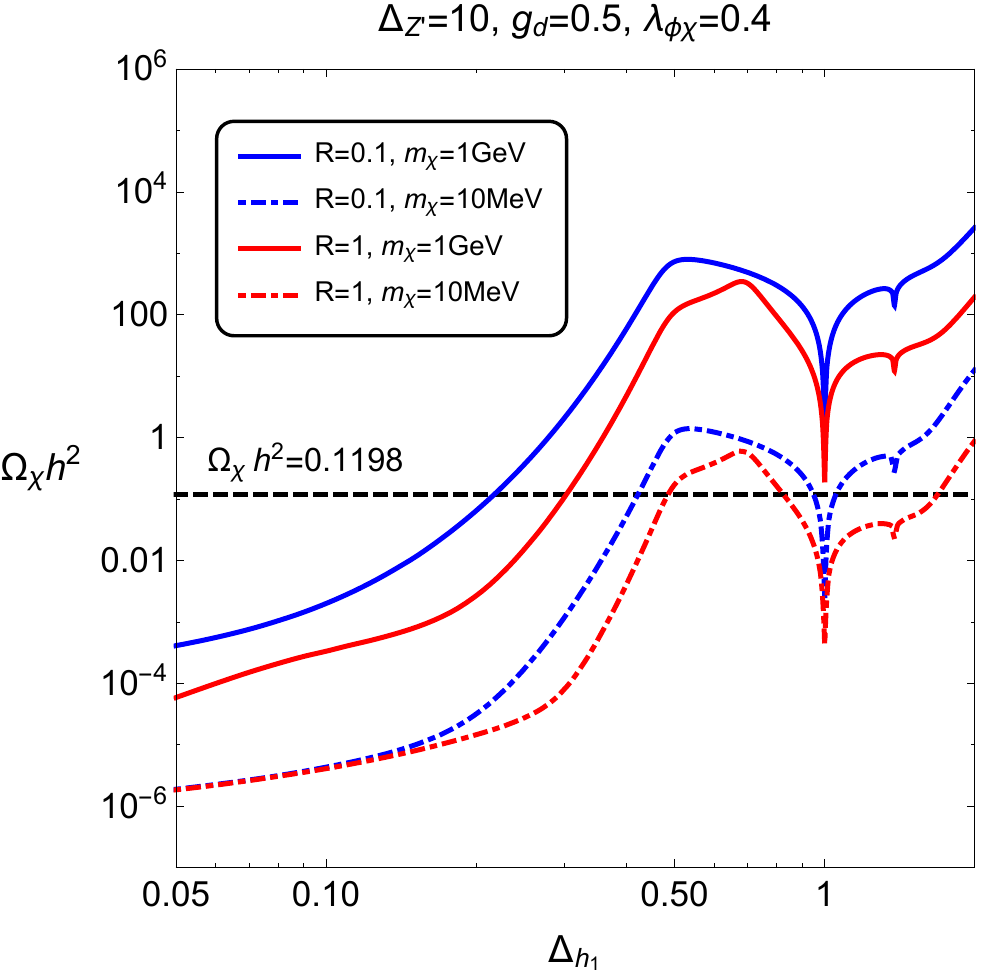}
   \end{center}
  \caption{Dark matter relic density as a function of  $\Delta_{h_1}$ for only forbidden channels with $h_1$ on left and both SIMP and forbidden channels with $h_1$ on right. 
 We have taken $(R,m_\chi)=(0.1, m_\chi=1\,{\rm GeV})$, $(0.1, m_\chi=10\,{\rm MeV})$, $(1.0, m_\chi=1\,{\rm GeV})$ and $(1.0, m_\chi=10\,{\rm MeV})$, from top to bottom. Black dashed lines correspond to the central value of relic density, $\Omega_\chi h^2=0.1198$, from Planck.   In both plots, we chose $g_d=0.5$, $\lambda_{\phi\chi}=0.4$ and $\Delta_{Z'}=10$.  }
  \label{forbh}
\end{figure}

Therefore, we can rewrite the Boltzmann equation by using the detailed balance conditions, (\ref{H1H1})  and (\ref{H1chi}), as follows,
\bea
\frac{dY_{\rm DM}}{dx}&=&-\lambda x^{-2}  \langle\sigma v\rangle_{h_1h_1\rightarrow \chi\chi^*} \left(\frac{1}{2}(1+\Delta_{h_1})^3 e^{-2\Delta_{h_1}x} \,Y^2_{\rm DM}-2(Y^{\rm eq}_{h_1})^2\right) \nonumber \\
&&-\lambda x^{-2}  \langle\sigma v\rangle_{h_1\chi^*\rightarrow \chi\chi}  \left(\frac{1}{2}(1+\Delta_{h_1})^{3/2} e^{-\Delta_{h_1} x} \,
Y^2_{\rm DM}- Y^{\rm eq}_{h_1} Y_{\rm DM}\right). \label{dH-Boltz}
\eea
As in the case with $Z'$ channels, the approximate solution to the above Boltzmann equation is then  given by
\bea
(Y_{\rm DM}(\infty))^{-1}
&=&\lambda \int^\infty_{x_f} dx \, x^{-2}\Big(\frac{1}{2}(1+\Delta_{h_1})^3 e^{-2\Delta_{h_1} x} \langle\sigma v\rangle_{h_1h_1\rightarrow \chi\chi^*} \nonumber \\
&& +   \frac{1}{2}(1+\Delta_{h_1})^{3/2} e^{-\Delta_{h_1} x}   \langle\sigma v\rangle_{h_1\chi^*\rightarrow \chi\chi}  \Big).
\eea

\begin{figure}
  \begin{center}
      \includegraphics[height=0.45\textwidth]{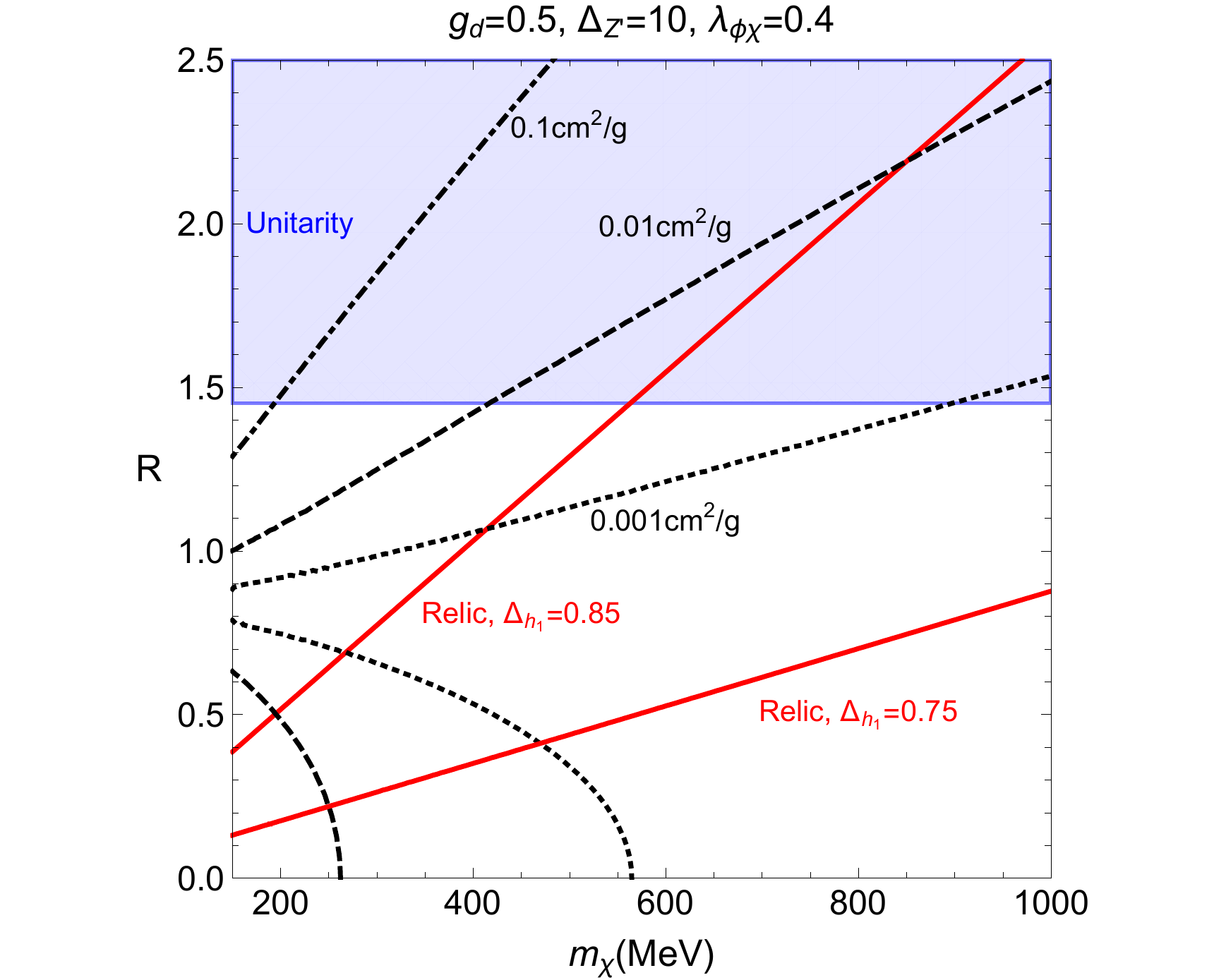}
   \end{center}
  \caption{Parameter space of $R$ vs $m_\chi$ for forbidden channels with $h_1$. The red lines satisfy the relic density and the blue region is excluded by unitarity. Dotted, dashed and dot-dashed lines correspond to self-scattering cross sections, $\sigma_{\rm self}/m_\chi=0.001, 0.01, 0.1\,{\rm cm^2/g}$. We took $\Delta_{h_1}=0.75$ or $ 0.85$ and the value of $\lambda_\chi$ saturates the vacuum stability bound, and $g_d=0.5$, $\lambda_{\phi\chi}=0.4$ and $\Delta_{Z'}=10$. }
  \label{forbh2}
\end{figure}

Expressing $ (\sigma v)_{h_1\chi^*\rightarrow \chi\chi}=c_1 $ and $(\sigma v)_{h_1 h_1\rightarrow \chi\chi^*} =c_2$, where $c_1$ and $c_2$ are given in eqs.~(\ref{h1h1}) and (\ref{h1chi}), we get the DM abundance as
\be
Y_{\rm DM}(\infty)\approx \frac{x_f}{\lambda}\, e^{\Delta_{h_1}x_f}\, h(\Delta_{h_1},x_f)
\ee
with
\bea
h(\Delta_{h_1},x_f)&=&\bigg[ \frac{1}{2} c_1 (1+\Delta_{h_1})^{3/2} \Big(1-\Delta_{h_1} x_f\, e^{\Delta_{h_1}x_f} \int^\infty_{\Delta_{h_1}x_f} dt\, t^{-1}\,e^{-t} \Big) \nonumber \\
&&+\frac{1}{2}c_2 (1+\Delta_{h_1})^3 e^{-\Delta_{h_1}x_f}  \Big(1-2\Delta_{h_1} x_f\, e^{2\Delta_{h_1}x_f}  \int^\infty_{2\Delta_{h_1}x_f} dt\, t^{-1}\,e^{-t} \Big)\bigg]^{-1}.
\eea
Consequently, the relic density is determined to be
\bea
\Omega_{\rm DM} h^2 
=5.20\times 10^{-10} {\rm GeV}^{-2} \Big(\frac{g_*}{10.75} \Big)^{-1/2}\Big(\frac{x_f}{20}\Big)\, e^{\Delta_{h_1} x_f}\, h(\Delta_{h_1}, x_f).
\eea

In Fig.~\ref{forbh}, we depicted the relic density as a function of $\Delta_{h_1}$, only with forbidden channels involving $h_1$ on left and with both SIMP an forbidden channels on right, varying $(R, m_\chi)$ such that $R=0.1-1$ and $m_\chi=10\,{\rm MeV}-1\,{\rm GeV}$. For both plots, we took $g_d=0.5$, $\lambda_{\phi\chi}=0.4$ and $\Delta_{Z'}=10$. 
 Inclusion of the SIMP processes on the right plot clearly shows a resonance behavior at $\Delta_{h_1}\sim 1$ or $m_{h_1}\sim 2m_\chi$, drastically changing the relic density to much smaller values. But, the same resonance also appears in the self-scattering of dark matter as can be seen in eq.~(\ref{chichistar}), so it would be in a tension with the bound from Bullet cluster.  Below the resonance region on the right plot, between $\Delta_{h_1}\sim 0.5-1$, there appears a similar plateau with a fixed relic density, that is dominantly determined by the SIMP processes.

In Fig.~\ref{forbh2}, we also show the parameter space of $m_\chi$ vs $R$, satisfying the relic density in red lines, for $\Delta_{h_1}=0.75$ and $0.85$, from bottom to top. We chose the value of $\lambda_\chi$ such that the vacuum stability bound is saturated and set $g_d=0.5$, $\lambda_{\phi\chi}=0.4$ and $\Delta_{Z'}=10$.  The blue region is excluded by unitarity and the contours with self-scattering cross section are shown similarly to those in Fig.~\ref{forbZ2}.

\subsection{The case with $m_\chi<m_{Z'}\sim m_{h_1}$}

\begin{figure}
  \begin{center}
   \includegraphics[height=0.42\textwidth]{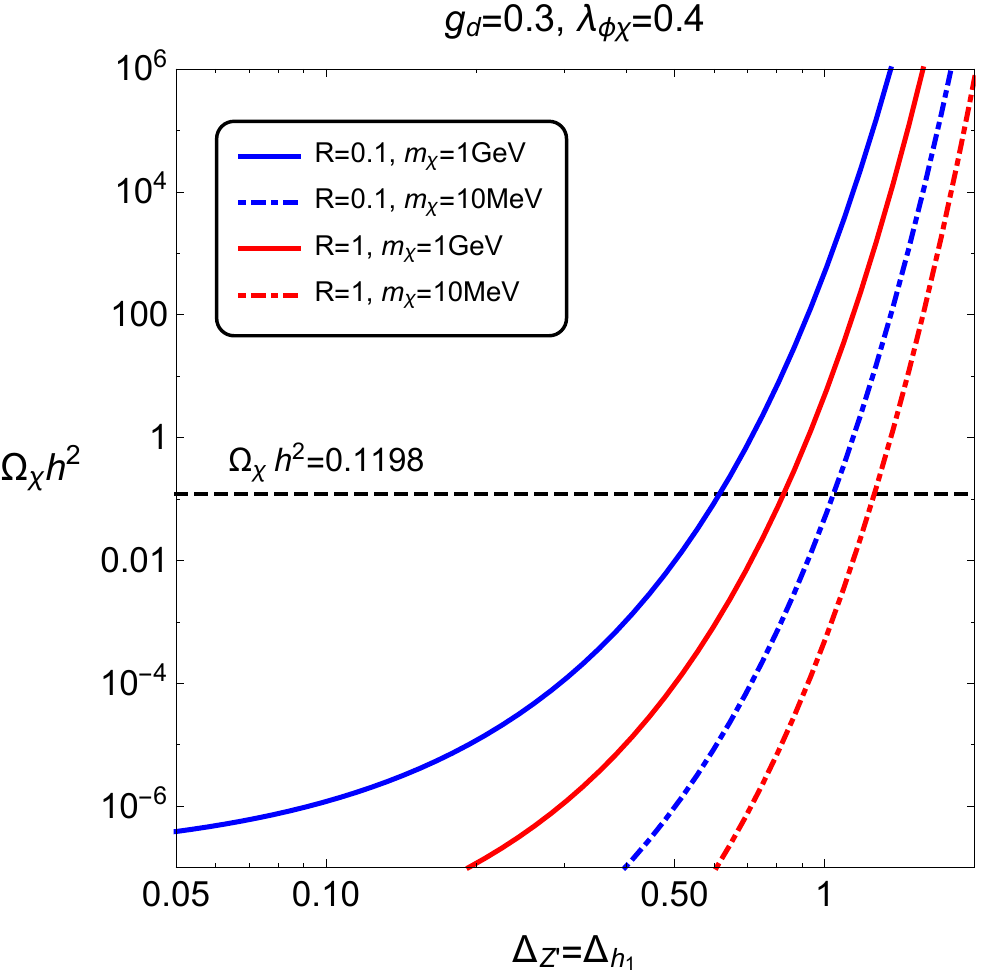}
      \includegraphics[height=0.42\textwidth]{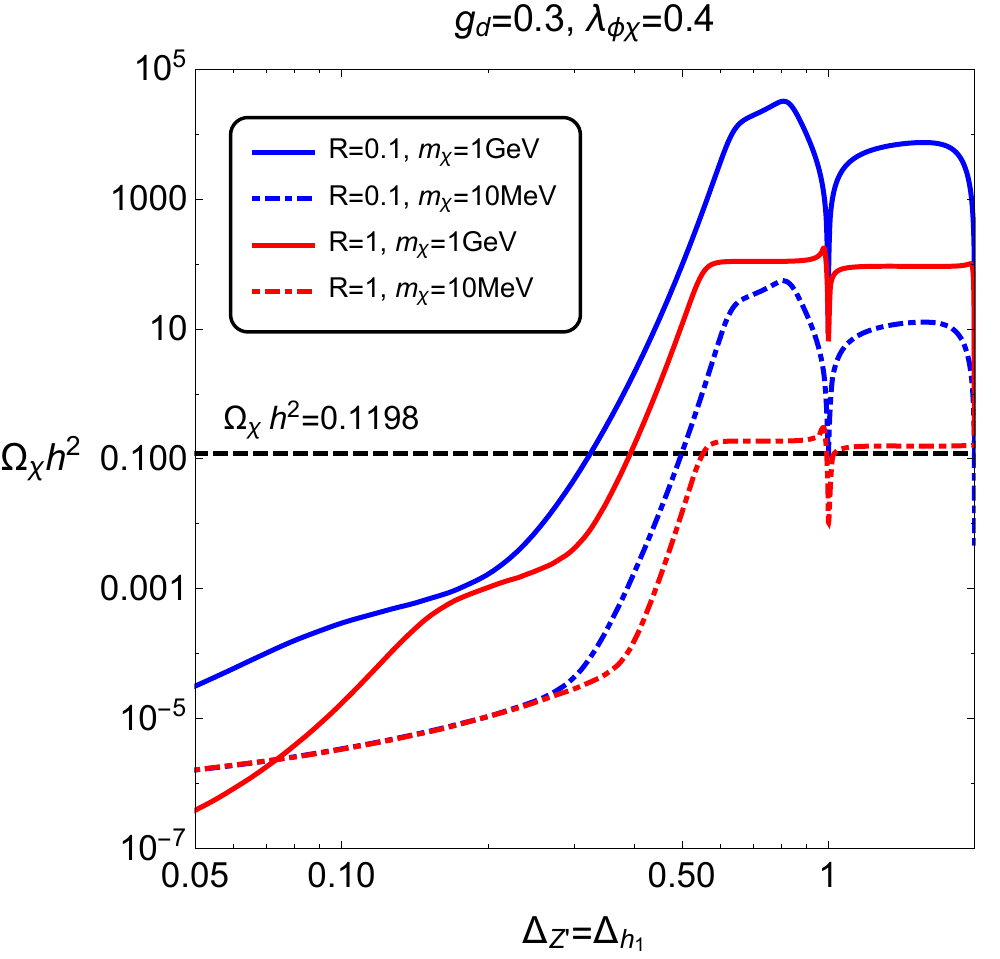}
   \end{center}
  \caption{Dark matter relic density as a function of  $\Delta_{h_1}=\Delta_{Z'}$ for only forbidden channels  with $Z'$ and $h_1$ on left and both SIMP and forbidden channels  with $Z'$ and $h_1$ on right. We have taken $(R,m_\chi)=(0.1, m_\chi=1\,{\rm GeV})$, $(0.1, m_\chi=10\,{\rm MeV})$, $(1.0, m_\chi=1\,{\rm GeV})$ and $(1.0, m_\chi=10\,{\rm MeV})$, from top to bottom. Black dashed lines correspond to the central value of relic density, $\Omega_\chi h^2=0.1198$, from Planck. 
In both plots, we chose $g_d=0.3$, $\lambda_{\phi\chi}=0.4$. }
  \label{forbboth}
\end{figure}

When $Z'$ and dark Higgs are comparably light, they both can contribute comparably to the forbidden channels at the same time. In this case, from eqs.~(\ref{Zp-Boltz}) and (\ref{dH-Boltz}), we obtain the approximate Boltzmann equation (\ref{genBoltz}) as
\bea
\frac{dY_{\rm DM}}{dx}&\approx&- \lambda x^{-2} \langle\sigma v\rangle_{\rm forb} Y^2_{\rm DM}
\eea
with
\bea
\langle\sigma v\rangle_{\rm forb}&=& \frac{9}{2}(1+\Delta_{Z'})^3 e^{-2\Delta_{Z'}x} \langle\sigma v\rangle_{Z'Z'\rightarrow \chi\chi^*} \nonumber \\
&& +   \frac{3}{2}(1+\Delta_{Z'})^{3/2} e^{-\Delta_{Z'} x}   \langle\sigma v\rangle_{Z'\chi^*\rightarrow \chi\chi} \nonumber \\
&&+\frac{1}{2}(1+\Delta_{h_1})^3 e^{-2\Delta_{h_1} x} \langle\sigma v\rangle_{h_1h_1\rightarrow \chi\chi^*} \nonumber \\
&& +   \frac{1}{2}(1+\Delta_{h_1})^{3/2} e^{-\Delta_{h_1} x}   \langle\sigma v\rangle_{h_1\chi^*\rightarrow \chi\chi}.  
\eea
Therefore, the DM  relic abundance becomes
\bea
Y_{\rm DM}(\infty)\approx \frac{x_f}{\lambda}\,\frac{e^{(\Delta_{Z'}+\Delta_{h_1})x_f/2} g\,h}{e^{(\Delta_{Z'}-\Delta_{h_1})x_f/2} g+ e^{-(\Delta_{Z'}-\Delta_{h_1})x_f/2} h}.
\eea
In this case, the relic density is given by
\bea
\Omega_{\rm DM} h^2 
=5.20\times 10^{-10} {\rm GeV}^{-2} \Big(\frac{g_*}{10.75} \Big)^{-1/2}\Big(\frac{x_f}{20}\Big)\, \frac{e^{(\Delta_{Z'}+\Delta_{h_1})x_f/2} g\,h}{e^{(\Delta_{Z'}-\Delta_{h_1})x_f/2} g+ e^{-(\Delta_{Z'}-\Delta_{h_1})x_f/2} h}.
\eea

In Fig.~\ref{forbboth}, we depicted the relic density as a function of $\Delta_{Z'}=\Delta_{h_1}$, when dark photon and dark Higgs are degenerate in mass, for varying $(R,m_\chi)$ between $R=0.1-1$ and $m_\chi=10\,{\rm MeV}-1\,{\rm GeV}$.  For both plots, we took $g_d=0.3$ and $\lambda_{\phi\chi}=0.4$. As in the case with light $Z'$ or $h_1$ in the previous subsections, there is a similar dependence on $\Delta_{Z'}$ as well as $(R,m_\chi)$.

\begin{figure}
  \begin{center}
      \includegraphics[height=0.45\textwidth]{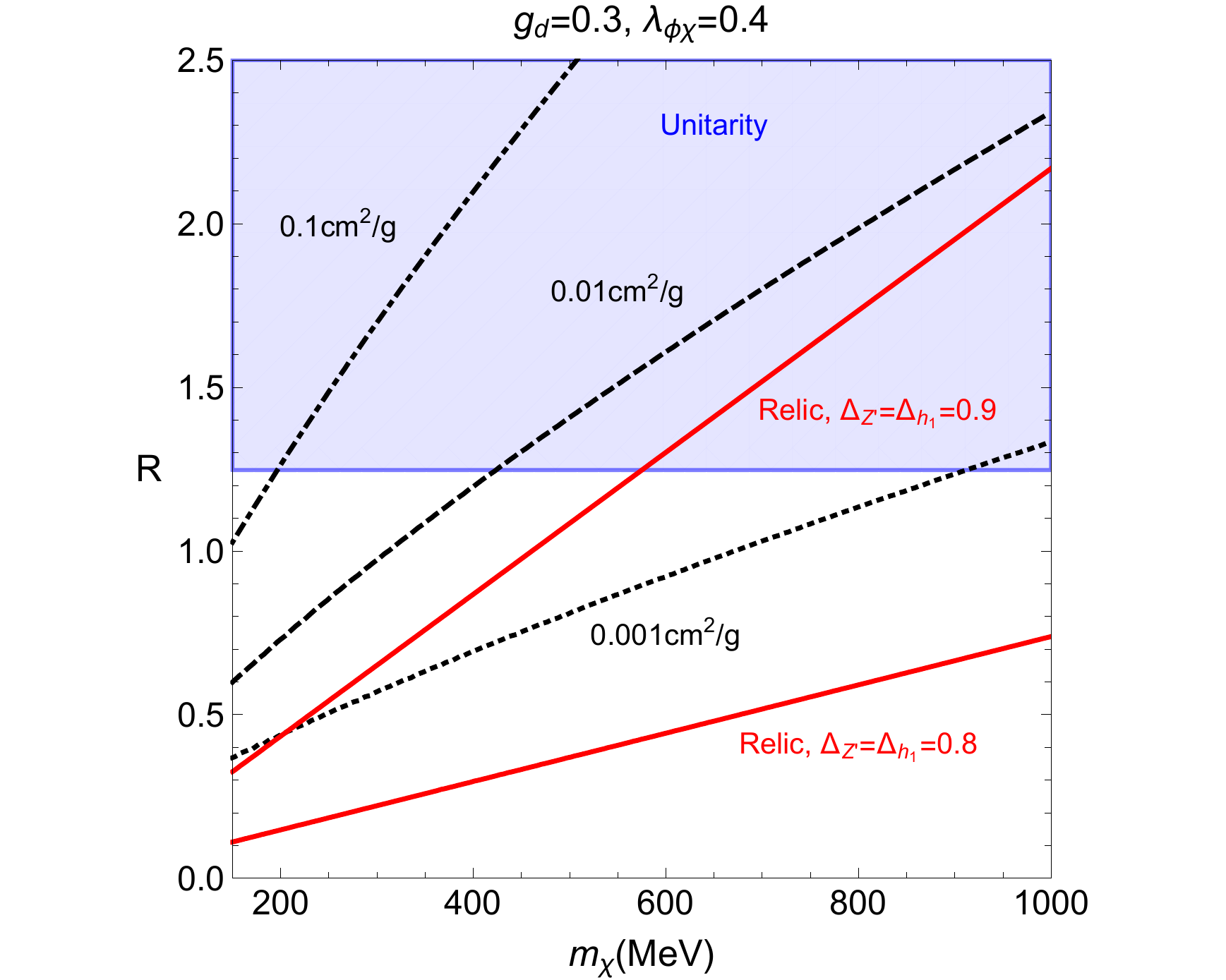}
   \end{center}
  \caption{Parameter space of $R$ vs $m_\chi$ for forbidden channels with $Z'$ and $h_1$. The red lines satisfy the relic density and the blue region is excluded by unitarity. Dotted, dashed and dot-dashed lines correspond to self-scattering cross sections, $\sigma_{\rm self}/m_\chi=0.001, 0.01, 0.1\,{\rm cm^2/g}$. We took $\Delta_{Z'}=\Delta_{h_1}=0.8$ or $0.9$ and the value of $\lambda_\chi$ saturates the vacuum stability bound, and  $g_d=0.3$, $\lambda_{\phi\chi}=0.4$. }
  \label{forbboth2}
\end{figure}

In Fig.~\ref{forbboth2}, we showed the parameter space for $m_\chi$ vs $R$, that explains the observed relic density in red lines, for $\Delta_{Z'}=\Delta_{h_1}=0.8$ and $0.9$, from bottom to top. 
We chose the value of $\lambda_\chi$ such that the vacuum stability bound is saturated and set $g_d=0.3$ and $\lambda_{\phi\chi}=0.4$.  The blue region is excluded by unitarity and the contours with self-scattering cross section are shown similarly to those in Fig.~\ref{forbZ2}.

\section{Conclusions}

We have considered the thermal production of self-interacting dark matter in models with $Z_3$ gauged symmetry. We showed that standard $2\rightarrow 2$ annihilation and hidden sector annihilations ($3\rightarrow 2$ annihilation and forbidden channels) can contribute equally in determining the relic density.  In particular, dark photon and dark Higgs in the model must be kept light for unitarity, so they both can contribute to the processes of dark matter annihilation. 
In particular, we found that forbidden channels with semi-annihilation such as $\chi\chi\rightarrow \chi^* Z'$ or $\chi\chi\rightarrow \chi^* h_1$ assist a thermal production of light dark matter with larger masses than in the SIMP case, but keeping a sizable self-scattering of dark matter.
Depending on the value of the self-scattering cross section favored by small-scale problems, we can identify the relevant thermal production mechanisms for self-interacting dark matter, in the same model.

\section*{Acknowledgments}

We would like to thank Kristjan Kannike for discussion on the vacuum stability condition. 
The work is supported in part by Basic Science Research Program through the National Research Foundation of Korea (NRF) funded by the Ministry of Education, Science and Technology (NRF-2016R1A2B4008759). The work of SMC is supported in part by TJ Park Science Fellowship of POSCO TJ Park Foundation.

\def\theequation{A.\arabic{equation}}

\setcounter{equation}{0}

\vskip0.8cm
\noindent
{\Large \bf Appendix A: $2\rightarrow 2$ annihilation for forbidden channels} 
\vskip0.4cm
\noindent

We summarize the formulas for annihilation cross sections that are relevant for forbidden channels.
The semi-annihilation involves dark photon or dark Higgs and it contributes only when the DM cubic coupling, $\kappa$, does not vanish. 

First, the cross sections for inverse processes of dark matter annihilation associated with $Z'$ are

$Z'Z'\rightarrow \chi \chi^*$:
\bea
(\sigma v)_{Z'Z'}&=& \frac{g_d^4}{72 \pi m_{Z'}^6 (m_{h_1}^2 - 4 m_{Z'}^2)^2}\bigg( (m_{h_1}^2 - 4 m_{Z'}^2)^2 (16 m_\chi^4 - 24 m_\chi^2 m_{Z'}^2 + 
      11 m_{Z'}^4)  \nonumber \\
      &&+ 18  \lambda_{\phi\chi} 
     m_{Z'}^2 (m_{h_1}^2 - 4 m_{Z'}^2) (-4 m_\chi^2 + m_{Z'}^2) v^2_d + 
   243  \lambda_{\phi\chi}^2 m_{Z'}^4 v_d^4\bigg) \sqrt{
 1 - \frac{m_\chi^2}{m_{Z'}^2}} \nonumber \\ \label{zpzp}
\eea

$Z'\chi^* \rightarrow \chi\chi$:
\bea
(\sigma v)_{Z'\chi^*}&=&\frac{g_d^2\kappa^2 v_d^2}{144\pi m_\chi^3 m_{Z'}^2} 
\frac{(3 m_\chi + m_{Z'})^2 }{ (m_\chi + m_{Z'})^3 (2 m_\chi + m_{Z'})^2} \nonumber \\
&&\times \Big(11 m_\chi^4 + 4 m_\chi^3 m_{Z'} -4 m_\chi^2 m_{Z'}^2 + m_{Z'}^4\Big)\sqrt{\Big(1 - \frac{m_\chi}{m_{Z'}}\Big) \Big(1 + \frac{3 m_\chi}{m_{Z'}}\Big) } \,v^2  \label{zpchi}
\eea

Secondly, the cross sections for inverse processes of dark matter annihilation associated with $h_1$ are

$h_1 h_1 \rightarrow \chi \chi^*$:
\bea
(\sigma v)_{h_1h_1}&=&\frac{\lambda_{\phi\chi}^2}{32 \pi m_{h_1}^2} 
\bigg(1 + \frac{
   2 v_d^2}{m_{h_1}^2} (\lambda_\phi-\lambda_{\phi\chi}) \bigg)^2\sqrt{1 - \frac{m_\chi^2}{m_{h_1}^2}}
   \label{h1h1} 
\eea

$h_1 \chi^*\rightarrow \chi\chi$:
\bea
(\sigma v)_{h_1 \chi^*}&=&\frac{\kappa^2}{64\pi m_\chi^3 m_{h_1}^2} 
\frac{\Big( m_\chi m_{h_1} (2 m_\chi + m_{h_1}) - 
\lambda_{\phi\chi} (3 m_\chi + 2 m_{h_1}) v_d^2\Big)^2}{
 (m_\chi + m_{h_1}) (2 m_\chi + m_{h_1})^2}\nonumber \\
 &&\times \sqrt{\Big(1 - \frac{m_\chi}{m_{h_1}}\Big) \Big(1 + \frac{3m_\chi}{m_{h_1}}\Big)}.  \label{h1chi}
\eea

\end{document}